\documentclass[preprint2]{proto}
\bibpunct{(}{)}{;}{a}{,}{,}
\usepackage{natbib}
\usepackage{times}

\usepackage{color}
\usepackage{amsbsy}
\usepackage{amsmath}

\newcommand{\sub}[1]{_{\mbox{\scriptsize #1}}}
\newcommand{\vc}[1]{{\boldsymbol #1}}

\begin{document}

\title{\textbf{\LARGE The Multifaceted Planetesimal Formation Process}}

\author {\textbf{\large Anders Johansen}}
\affil{\small\em Lund University}

\author {\textbf{\large J\"urgen Blum}}
\affil{\small\em Technische Universit\"at Braunschweig}

\author {\textbf{\large Hidekazu Tanaka}}
\affil{\small\em Hokkaido University}

\author {\textbf{\large Chris Ormel}}
\affil{\small\em University of California, Berkeley}

\author {\textbf{\large Martin Bizzarro}}
\affil{\small\em Copenhagen University}

\author {\textbf{\large Hans Rickman}}
\affil{\small\em Uppsala University\\Polish Academy of Sciences Space Research
Center, Warsaw}

\begin{abstract}
\baselineskip = 11pt
\leftskip = 0.65in 
\rightskip = 0.65in
\parindent=1pc
{\small Accumulation of dust and ice particles into planetesimals is an
important step in the planet formation process. Planetesimals are the seeds of
both terrestrial planets and the solid cores of gas and ice giants forming by
core accretion. Left-over planetesimals in the form of asteroids,
trans-Neptunian objects and comets provide a unique record of the physical
conditions in the solar nebula. Debris from planetesimal collisions around
other stars signposts that the planetesimal formation process, and hence planet
formation, is ubiquitous in the Galaxy. The planetesimal formation stage
extends from micrometer-sized dust and ice to bodies which can undergo run-away
accretion. The latter ranges in size from 1 km to 1000 km, dependent on the
planetesimal eccentricity excited by turbulent gas density fluctuations.
Particles face many barriers during this growth, arising mainly from
inefficient sticking, fragmentation and radial drift. Two promising growth
pathways are mass transfer, where small aggregates transfer up to 50\% of their
mass in high-speed collisions with much larger targets, and fluffy growth,
where aggregate cross sections and sticking probabilities are enhanced by a low
internal density. A wide range of particle sizes, from mm to 10 m, concentrate
in the turbulent gas flow. Overdense filaments fragment gravitationally into
bound particle clumps, with most mass entering planetesimals of contracted
radii from 100 to 500 km, depending on local disc properties. We propose a
hybrid model for planetesimal formation where particle growth starts unaided by
self-gravity but later proceeds inside gravitationally collapsing pebble clumps
to form planetesimals with a wide range of sizes.
 \\~\\~\\~}

\end{abstract}  

\section{\textbf{INTRODUCTION}}

Most stars are born surrounded by a thin protoplanetary disc with a
characteristic mass between 0.01\% and 10\% of the mass of the central star
\citep{AndrewsWilliams2005}. Planetesimal formation takes place as the embedded
dust and ice particles collide and grow to ever larger bodies. Tiny particles
collide gently due to Brownian motion, while larger aggregates achieve higher
and higher collision speeds as they gradually decouple from the smallest eddies
of the turbulent gas flow and thus no longer inherit the incompressibility of
the gas \citep{Voelk+etal1980}. The gas disc is partially
pressure-supported in the radial direction, causing particles of sizes from
centimeters to meters to drift towards the star
\citep{Whipple1972,Weidenschilling1977a}. Drift speeds depend on the particle
size and hence different-sized particles experience high-speed collisions.

The growth from dust and ice grains to planetesimals -- the seeds of both
terrestrial planets as well as the cores of gas giants and ice giants -- is an
important step towards the assembly of a planetary system. A planetesimal can
be defined as a body which is held together by self-gravity rather than
material strength, corresponding to minimum sizes of order 100--1,000 meters
\citep{Benz2000}. The planetesimal formation {\it stage}, on the other hand,
must extend to sizes where the escape speed of the largest bodies exceeds the
random motion of the planetesimals, to enter the next stage of gravity-driven
collisions. The random speed of planetesimals and preplanetesimals (the latter
can be roughly defined as bodies larger than 10 meters in size) is excited
mainly by the stochastic gravitational pull from density fluctuations in the
turbulent gas; under nominal turbulence conditions the largest planetesimals
need to reach sizes as large as 1,000 km to start run-away growth
\citep{Ida+etal2008}. In disc regions with less vigorous turbulence,
planetesimal sizes between 1 and 10 kilometers may suffice to achieve
significant gravitational focusing \citep{Gressel+etal2012}.

Planetesimals must grow to run-away sizes despite bouncing and disruptive
collisions \citep{BlumWurm2000}. Planetesimals must also grow rapidly -- radial
drift time-scales of cm-m-sized particles are as short as a few hundred
orbits. Despite these difficulties there is ample evidence from cosmochemistry
that large planetesimals formed in the solar nebula within a few million years
\citep{Kleine+etal2004,Baker+etal2005}, early enough to melt and differentiate
by decay of short-lived radionuclides.

The time since the last {\em Protostars \& Planets} review on planetesimal
formation \citep{Dominik+etal2007} has seen a large expansion in the complexity
and realism of planetesimal formation studies. In this review we focus on three
areas in which major progress has been obtained, namely (i) the identification
of a bouncing barrier at mm sizes and the possibility of growth by mass
transfer in high-speed collisions
\citep{Wurm+etal2005,Johansen+etal2008,Guettler+etal2010,Zsom+etal2010,Windmark+etal2012a},
(ii) numerical simulations of collisions between highly porous ice aggregates
which can grow past the bouncing and radial drift barriers due to efficient
sticking and cross sections that are greatly enhanced by a low internal density
\citep{Wada+etal2008,Wada+etal2009,SeizingerKley2013}, and (iii) hydrodynamical
and magnetohydrodynamical simulations which have identified a number of
mechanisms for concentrating particles in the turbulent gas flow of
protoplanetary discs, followed by a gravitational fragmentation of the
overdense filaments to form planetesimals with characteristic radii larger than
100 km
\citep{Johansen+etal2007,Lyra+etal2008b,Lyra+etal2009a,Johansen+etal2009b,BaiStone2010b,Raettig+etal2013}.

The review is laid out as follows. In section \ref{s:belts} we give an overview
of the observed properties of planetesimal belts around the Sun and other
stars. In section \ref{s:formation} we review the physics of planetesimal
formation and derive the planetesimal sizes necessary to progress towards
planetary systems. Section \ref{s:concentration} concerns particle
concentration and the density environment in which planetesimals form, while
the following section \ref{s:dustgrowth} describes laboratory experiments to
determine the outcome of collisions and methods for solving the coagulation
equation. In section \ref{s:coagulation-fragmentation} we discuss the bouncing
barrier for silicate dust and the possibility of mass transfer in high-speed
collisions. Section \ref{s:fluffy} discusses computer simulations of highly
porous ice aggregates whose low internal density can lead to high growth
rates. Finally in section \ref{s:unified} we integrate the knowledge gained
over recent years to propose a unified model for planetesimal formation
involving both coagulation, particle concentration and self-gravity.

\section{\textbf{PLANETESIMAL BELTS}}
\label{s:belts}

The properties of observed planetesimal belts provide important constraints on
the planetesimal formation process. In this section we review the main
properties of the asteroid belt and trans-Neptunian population as well as
debris discs around other stars.

\subsection{Asteroids}

The asteroid belt is a collection of mainly rocky bodies residing between the
orbits of Mars and Jupiter. Asteroid orbits have high eccentricities
($e$$\sim$$0.1$) and inclinations, excited by resonances with Jupiter and also
a speculated population of embedded super-Ceres-sized planetary embryos which
were later dynamically depleted from the asteroid belt \citep{Wetherill1992}.
The high relative speeds imply that the asteroid belt is in a highly erosive
regime where collisions lead to fragmentation rather than to growth.

Asteroids range in diameters from $D\approx1000$ km (Ceres) down to the
detection limit at sub-km sizes \citep{Gladman+etal2009}. The asteroid size
distribution can be parameterised in terms of the cumulative size distribution
$N_>(D)$ or the differential size distribution ${\rm d}N_>/{\rm d} D$. The
cumulative size distribution of the largest asteroids resembles a broken power
law $N_> \propto D^{-p}$, with a steep power law index $p\approx3.5$ for
asteroids above 100 km diameter and a shallower power law index $p\approx1.8$
below this knee \citep{Bottke+etal2005}.

Asteroids are divided into a number of classes based on their spectral
reflectivity. The main classes are the moderate-albedo S-types which dominate
the region from 2.1 to 2.5 AU and the low-albedo C-types which dominate regions
from 2.5 AU to 3.2 AU \citep{GradieTedesco1982,BusBinzel2002}. If these classes
represent distinct formation events, then their spatial separation can be used
to constrain the degree of radial mixing by torques from turbulent density
fluctuations in the solar nebula \citep{NelsonGressel2010}. Another important
class of asteroids is the M-type of which some are believed to be the metallic
cores remaining from differentiated asteroids \citep{Rivkin+etal2000}. Two of
the largest asteroids in the asteroid belt, Ceres and Vesta, are known to be
differentiated \citep[from the shape and measured gravitational moments,
respectively,][]{Thomas+etal2005,Russell+etal2012}.

Asteroids are remnants of solar system planetesimals that have undergone
substantial dynamical depletion and collisional erosion. Dynamical evolution
models can be used to link their current size distribution to the primordial
birth size distribution. \cite{Bottke+etal2005} suggested, based on the
observed knee in the size distribution at sizes around 100 km, that asteroids
with diameters above $D\simeq120$ km are primordial and that their steep size
distribution reflects their formation sizes, while smaller asteroids are
fragments of collisions between their larger counterparts.
\cite{Morbidelli+etal2009} tested a number of birth size distributions for the
asteroid belt, based on either classical coagulation models starting with
km-sized planetesimals or turbulent concentration models predicting birth sizes
in the range between 100 and 1000 km \citep{Johansen+etal2007,Cuzzi+etal2008},
and confirmed that the best match to the current size distribution is that the
asteroids formed big, in that the asteroids above 120 km in diameter are
depleted dynamically but have maintained the shape of the primordial size
distribution. On the other hand, \cite{Weidenschilling2011} found that an
initial population of 100-meter-sized planetesimals can reproduce the current
observed size distribution of the asteroids, including the knee at 100 km.
However, gap formation around large planetesimals and trapping of small
planetesimals in resonances is not included in this particle-in-a-box approach
\citep{Levison+etal2010}.

\subsection{Meteorites}
\label{s:meteorites}

Direct information about the earliest stages of planet formation in the solar
nebula can be obtained from meteorites that record the interior structure of
planetesimals in the asteroid belt. The reader is referred to the chapter by
{\it Davis et al.} for more details on the connection between cosmochemistry
and planet formation. Meteorites are broadly characterized as either primitive
or differentiated \citep{Krot+etal2003}. Primitive meteorites (chondrites) are
fragments of parent bodies that did not undergo melting and differentiation
and, therefore, contain pristine samples of the early solar system. In
contrast, differentiated meteorites are fragments of parent bodies that
underwent planetesimal-scale melting and differentiation to form a core, mantle
and crust.

\begin{figure}[!t]
  \epsscale{1.0}
  \includegraphics[width=\linewidth]{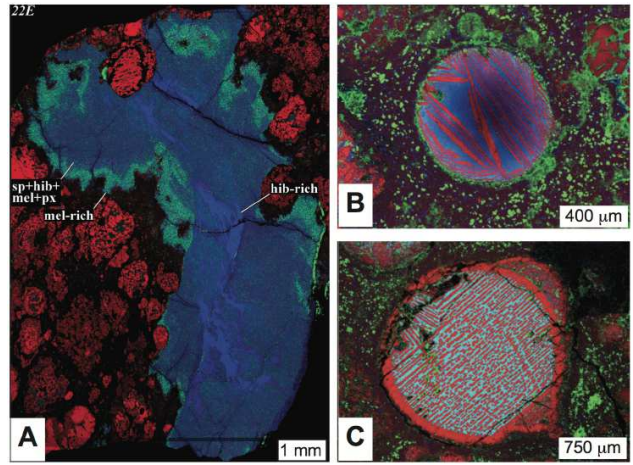}
  \caption{\small X-ray elemental maps composed of Mg (red), Ca (green), and Al
  (blue) of a fined-grained CAI from the Efremovka CV3 chondrite (A) and two
  barred-olivine chondrules (B, C) from the NWA 3118 CV3 chondrite.}
  \label{f:chondrules}
\end{figure}

The oldest material to crystallise in the solar nebula is represented in
chondrites as mm- to cm-sized calcium-aluminium-rich inclusions (CAIs) and
ferromagnesian silicate spherules (chondrules) of typical sizes from 0.1 mm to
1 mm (Fig.\ \ref{f:chondrules}). The chondrules originated in an unidentified
thermal processing mechanism (or several mechanisms) which melted pre-existing
nebula solids \citep[e.g.][]{DeschConnolly2002,Ciesla+etal2004}; alternatively
chondrules can arise from the crystallisation of splash ejecta from
planetesimal collisions \citep{SandersTaylor2005,Krot+etal2005}.

The majority of CAIs formed as fine-grained condensates from an $^{16}$O-rich
gas of approximately solar composition in a region with high ambient
temperature ($>$1300 K) and low total pressures ($\sim$$10^{-4}$ bar). This
environment may have existed in the innermost part of the solar nebula during
the early stage of its evolution characterized by high mass accretion rates
($\sim$$10^{-5}$ $M_\odot$ yr$^{-1}$) to the proto-Sun
\citep{D'Alessio+etal2005}. Formation of CAIs near the proto-Sun is also
indicated by the presence in these objects of short-lived radionuclide
$^{10}$Be formed by solar energetic particle irradiation
\citep{McKeegan+etal2000}. In contrast, most chondrules represent coalesced
$^{16}$O-poor dust aggregates that were subsequently rapidly melted and cooled
in lower-temperature regions ($<$1000 K) and higher ambient vapor pressures
($\ge10^{-3}$ bar) than CAIs, resulting in igneous porphyritic textures we
observe today ({\em Scott}, 2007). Judging by their sheer abundance in
chondrite meteorites, the formation of chondrules must reflect a common process
that operated in the early solar system. Using the assumption-free
uranium-corrected Pb-Pb dating method, \cite{Connelly+etal2012} recently showed
that CAIs define a brief formation interval corresponding to an age of 4567.30
$\pm$ 0.16 Myr, whereas chondrule ages range from 4567.32 $\pm$ 0.42 to 4564.71
$\pm$ 0.30 Myr. These data indicate that chondrule formation started
contemporaneously with CAIs and lasted $\sim$3 Myr.

A consequence of accretion of planetesimals on Myr time-scales or less is the
incorporation of heat-generating short-lived radioisotopes such as $^{26}$Al,
resulting in wide-scale melting, differentiation and extensive volcanic
activity. Both long-lived and short-lived radioisotope chronometers have been
applied to study the timescale of asteroidal differentiation. Of particular
interest are the $^{26}$Al-$^{26}$Mg and $^{182}$Hf-$^{182}$W decay systems,
with half-lives of 0.73 Myr and 9 Myr, respectively. Al and Mg are refractory
and lithophile elements and thus remain together in the mantle after core
formation. In contrast, the different geochemical behavior of Hf and W during
melting results in W being preferentially partitioned in the core and Hf being
partitioned into the silicate mantle and crust. Therefore, the short-lived
$^{182}$Hf-$^{182}$W system is useful to study the timescales of core formation
in asteroids, as well as in planets, which can be used to constrain
planetesimal formation models.

\begin{figure}[!t]
  \epsscale{1.0}
  \plotone{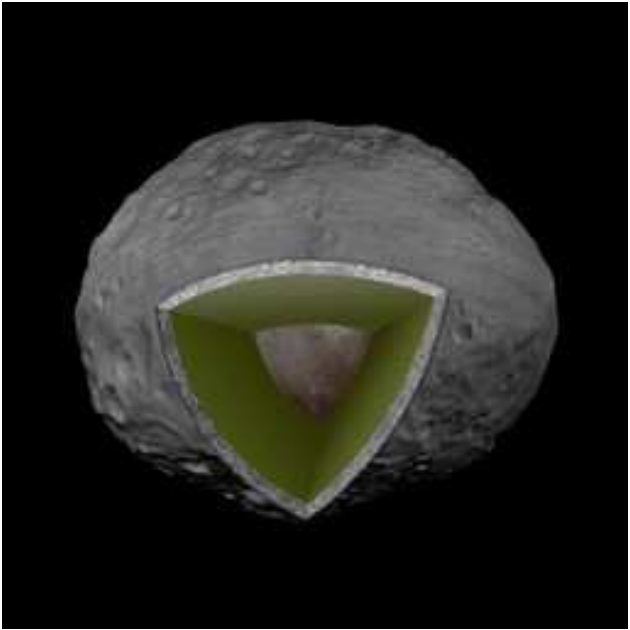}
  \caption{\small Surface and interior structure of the 500-km-diameter
  asteroid 4 Vesta, obtained by the NASA Dawn satellite. The iron core is 110
  km in radius, surrounded by a silicate mantle (green) and a basaltic crust
  (gray). Image credit: NASA/JPL-Caltech}
  \label{f:vesta}
\end{figure}

Eucrite and angrite meteorites are the two most common groups of basaltic
meteorites, believed to be derived from the mantles of differentiated parent
bodies. The HED (howardite-eucrite-diogenite) meteorite clan provides our best
samples of any differentiated asteroid and could come from the 500-km-diameter
asteroid 4 Vesta \citep{BinzelXu1993}, although see \cite{Schiller+etal2011}
for a different view. The interior structure of Vesta was recently determined
by the Dawn mission \citep{Russell+etal2012}, indicating an iron core of
approximately 110 km in radius (see Fig.\ \ref{f:vesta}). The
$^{26}$Al-$^{26}$Mg systematics of the angrite and HED meteorites indicate that
silicate differentiation on these parent bodies occurred within the first few
Myr of solar system formation
\citep{Schiller+etal2011,Bizzarro+etal2005,Schiller+etal2010,Spivak-Birndorf+etal2009}.
Similarly, the Mg isotope composition of olivines within pallasite meteorites
–- a type of stony-iron meteorites composed of cm-sized olivine crystals set in
a iron-nickel matrix –- suggests silicate differentiation within 1.5 Myr of
solar system formation \citep{Baker+etal2012}. Chemical and isotopic diversity
of iron meteorites show that these have sampled approximately 75 distinct
parent bodies \citep{Goldstein+etal2009}. The $^{182}$Hf-$^{182}$W systematics
of some magmatic iron meteorites require the accretion and differentiation of
their parent bodies to have occurred within less than 1 Myr of solar system
formation \citep{Kleine+etal2009}.

An important implication of the revised absolute chronology of chondrule
formation of \cite{Connelly+etal2012}, which is not based on short-lived
radionuclides for which homogeneity in the solar nebula has to be assumed, is
that the accretion and differentiation of planetesimals occurred during the
epoch of chondrule formation. This suggests that, similarly to chondrite
meteorites, the main original constituents of early-accreted asteroids may have
been chondrules. The timing of accretion of chondritic parent bodies can be
constrained by the ages of their youngest chondrules, which requires that most
chondrite parent bodies accreted $>$2 Myr after solar system formation.
Therefore, chondrites may represent samples of asteroidal bodies that formed
after the accretion of differentiated asteroids, at a time when the levels of
$^{26}$Al were low enough to prevent significant heating and melting. 

\subsection{Trans-Neptunian objects}

The trans-Neptunian objects are a collection of rocky/icy objects beyond the
orbit of Neptune \citep{LuuJewitt2002}. Trans-Neptunian objects are categorised
into several dynamical classes, the most important for planetesimal formation
being the classical Kuiper belt objects, the scattered disc objects and the
related centaurs which have orbits that cross the orbits of one or more of the
giant planets. The classical Kuiper belt objects do not approach Neptune at any
point in their orbits and could represent the pristine population of icy
planetesimals in the outer solar nebula. The so-called cold component of the
classical Kuiper belt has a large fraction (at least 30\%) of binaries of
similar size \citep{Noll+etal2008} which strongly limits the amount of
collisional grinding that these bodies can have undergone
\citep{Nesvorny+etal2011}. The scattered disc objects on the other hand can
have large semi-major axes, but they all have perihelia close to Neptune's
orbit. The centaurs move between the giant planets and are believed to be the
source of short period comets (see below).

The largest trans-Neptunian objects are much larger than the largest asteroids,
with Pluto, Haumea, Makemake and Eris defined as dwarf planets of 1.5-3 times
the diameter of Ceres \citep{Brown2008}. Nevertheless the size distribution of
trans-Neptunian objects shows similarities with the asteroid belt, with a steep
power law above a knee at around $D$$\sim$$100$ km \citep{FuentesHolman2008}.
The turn-over at the knee implies that there are fewer intermediate-mass
planetesimals than expected from an extrapolation from larger sizes -- this has
been dubbed the ``Missing Intermediate-Sized Planetesimals'' problem
\citep{SheppardTrujillo2010,Shankman+etal2013} and suggests that the
characteristic planetesimal birth size was $\sim$100 km. The accretion ages of
Kuiper belt objects are not known, in contrast to the differentiated asteroids
where the inclusion of large amounts of $^{26}$Al requires early accretion.
While the largest Kuiper belt objects are likely differentiated into a rocky
core and an icy mantle -- this is clear e.g.\ for Haumea which is the dense
remnant of a differentiated body \citep{RagozzineBrown2009} -- differentiation
could be due to long-lived radionuclides such as U, Th and $^{40}$K and hence
happen over much longer time-scales ($\sim$100 Myr) than the $\sim$Myr
time-scale characteristic of asteroid differentiation by $^{26}$Al decay
\citep{McKinnon+etal2008}.

\subsection{Comets}

Comets are typically km-sized volatile-rich bodies which enter the inner solar
system. They bring with them a wealth of information about the conditions
during the planetesimal formation epoch in the outer solar system. Outgassing
provides knowledge about the compositions and heating histories of icy
planetesimals, and the volumes and masses of comet nuclei can be used to derive
their density which can be compared to the density expected from planetesimal
formation models. Comet nuclei are (or are related to) icy planetesimals, i.e.\
planetesimals that formed beyond the snow line in a formation zone extending
from roughly 15 to 30 AU from the Sun -- at least in the framework of the
so-called Nice Model where the giant planets form in a compact configuration
between 5 and 12 AU and later migrate by planetesimal scattering to their
current positions \citep{Levison+etal2011}. This zone is wide enough to allow
for the appearance of chemical zoning, but such a zoning would likely not be
reflected in any separation between the short- and long-period comet source
regions.

Comets have to be considered together with the Centaurs and trans-Neptunians,
because these are dynamically related to the Jupiter Family comets, and most of
the data we have on comets come from observations and modeling of Jupiter
Family comets. The emergent picture is a very wide size spectrum going from
sub-km to $10^3$ km in diameter. However, the slope of the size distribution is
difficult to establish, since small objects are unobservable in the outer
populations, and large ones are lacking in the Jupiter Family. Generally, the
measurements of masses and sizes of comet nuclei are consistent with a rather
narrow range of densities at about 0.5~g/cm$^3$
\citep{Weissman+etal2004,Davidsson+etal2007}. Most of the determinations use
the non-gravitational force due to asymmetric outgassing. Assuming a
composition that is roughly a 50-50 mix of ice and refractories
\citep{GreenbergHage1990} the porosity comes out as roughly 2/3. But the mass
determinations only concern the bulk mass, so one cannot distinguish between
meso-scale-porosity ({\it rubble piles}) and small-scale porosity ({\it pebble
piles}). Estimates of comet tensile strengths have generally been extremely
low, as expected for porous bodies. For comet Shoemaker-Levy~9, modeling of its
tidal breakup led to (non-zero) values so low that the object was described as
a strengthless rubble pile \citep{AsphaugBenz1996}. The non-tidal splittings
often observed for other comets appear to be so gentle that, again, an
essentially zero strength has been inferred \citep{Sekanina1997}.

The current results on volatile composition of comets have not indicated any
difference between the Jupiter Family comets -- thought to probe the scattered
disc -- and the Halley-type and long-period comets, which should come from the
Oort Cloud \citep{A'Hearn+etal2012}. From a dynamical point of view, this
result is rather expected, since both these source populations have likely
emerged from the same ultimate source, a disc of icy planetesimals extending
beyond the giant planet zone in the early Solar System
\citep{BrasserMorbidelli2013}.

The origin of comets is closely related to the issue of the chemical
pristineness of comets. If the total mass of the planetesimal disc was
dominated by the largest planetesimals, several hundreds of km in size, then
the km-sized comet nuclei that we are familiar with could arise from
collisional grinding of the large planetesimals. Nevertheless there are severe
problems with this idea. One is that comet nuclei contain extremely volatile
species, like the S$_2$ molecule \citep{Bockelee-Morvan+etal2004}, which would
hardly survive the heating caused by a disruptive impact on the parent body.
The alternative that tidal disruptions of large planetesimals at mutual close
encounters may lead to large numbers of smaller objects seems more viable, but
this too suffers from the second pristineness problem, namely, the geologic
evolution expected within a large planetesimal due to short-lived radio nuclei
\citep{Prialnik+etal2004}. If the large planetesimal thus becomes chemically
differentiated, there seems to be no way to form the observed comet nuclei by
breaking it up -- no matter which mechanism we invoke.

The pristine nature of comets is thus consistent with the formation of small,
km-scale cometesimals in the outer part of the solar nebula, avoiding due to
their small size melting and differentiation due to release of short-lived
radionuclides. Another possibility is that comets accreted from material
comprising an early formed $^{26}$Al-free component \citep{Olsen+etal2013},
which would have prevented differentiation of their parent bodies. The recent
chronology of $^{26}$Al-poor inclusions found in primitive meteorites indicate
that this class of objects formed coevally with CAIs \citep[which record the
canonical $^{26}$Al/$^{27}$Al value of ∼$5 \times
10^{-5}$,][]{Holst+etal2013}. This provides evidence for the existence of
$^{26}$Al-free material during the earliest stages of the protoplanetry disc.
The discovery of refractory material akin to CAIs in the Stardust samples
collected from Comet 81P/Wild 2 shows efficient outward transport of material
from the hot inner disc regions to cooler environments far from the Sun during
the epoch of CAI formation \citep{Brownlee+etal2006}. Efficient outward
transport during the earliest stages of solar system evolution would have
resulted in the delivery of a significant fraction of the $^{26}$Al-poor
material to the accretion region of cometary bodies. Analysis of the Coci
refractory particle returned from comet 81P/Wild 2 did not show detectable
$^{26}$Al at the time of its crystallization \citep{Matzel+etal2010} which,
although speculative, is consistent with the presence an early-formed
$^{26}$Al-free component in comets.

\subsection{Debris discs}

Planetesimal belts around other stars show their presence through the infrared
emission of the dust produced in collisions \citep{Wyatt2008}. The spectral
energy distribution of the dust emission reveals the orbital distance of the
planetesimal belt. Warm debris discs resembling the asteroid belt in the Solar
System are common around young stars of around 50 Myr age (at least 30\%), but
their occurrence falls to a few percent within 100--1,000 Myr
\citep{Siegler+etal2007}. Thus planetesimal formation appears to be ubiquitous
around the Sun as well as around other stars. This is in agreement with results
from the Kepler mission that flat planetary systems exist around a high
fraction of solar-type stars
\citep{Lissauer+etal2011,TremaineDong2012,Johansen+etal2012b,FangMargot2012}.
\begin{figure*}[!t]
  \epsscale{1.7}
  \plotone{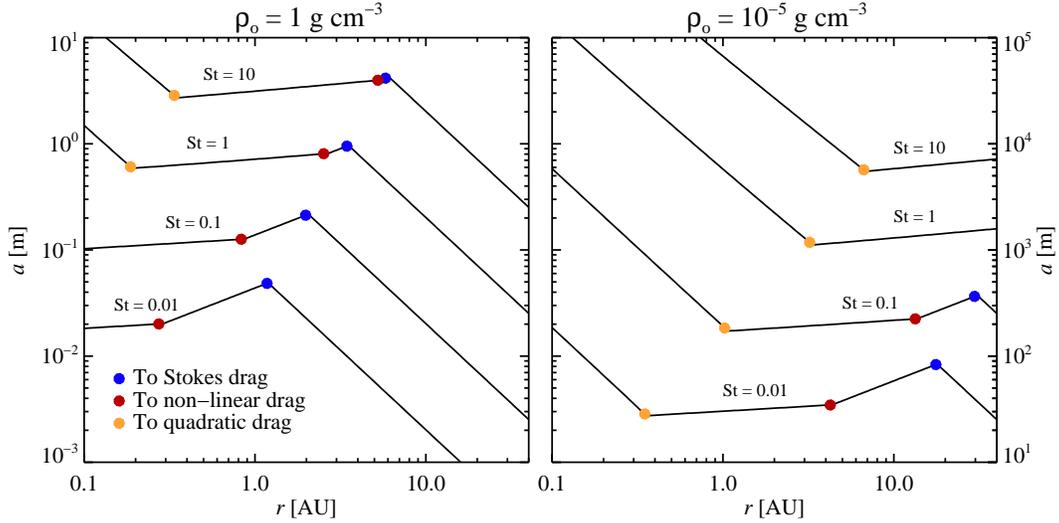}
  \caption{\small The particle size corresponding to Stokes numbers from $0.01$
  to $10$, the left plot for compact particles with material density
  $\rho_\bullet=1\,{\rm g\,cm^{-3}}$ and the right for extremely porous
  aggregates with $\rho_\bullet=10^{-5}\,{\rm g\,cm^{-3}}$. Such fluffy
  particles may be the result of collisions between ice aggregates (discussed
  in section \ref{s:fluffy}). Transitions between drag force regimes are
  indicated with large dots.}
  \label{f:a_r_st}
\end{figure*}

\section{\textbf{THE PLANETESIMAL FORMATION STAGE}}
\label{s:formation}

The dust and ice particles embedded in the gas in protoplanetary discs collide
and merge, first by contact forces and later by gravity. This process leads
eventually to the formation of the terrestrial planets and the cores of gas
giants and ice giants forming by core accretion. The planetesimal formation
stage can broadly be defined as the growth from dust grains to particle sizes
where gravity contributes significantly to the collision cross section of two
colliding bodies.

\subsection{Drag force}
\label{s:drag_force}

Small particles are coupled to the gas via drag force. The acceleration by the
drag force can be written as
\begin{equation}
  \dot{\vc{v}} = -\frac{1}{\tau_{\rm f}}
  (\vc{v}-\vc{u}) \, ,
\end{equation}
where $\vc{v}$ is the particle velocity, $\vc{u}$ is the gas velocity at the
position of the particle and $\tau_{\rm f}$ is the friction time which contains
all the physics of the interaction of the particle with the gas flow
\citep{Whipple1972,Weidenschilling1977a}. The friction time can be divided into
different regimes, depending on the mean free path of the gas molecules,
$\lambda$, and the speed of the particle relative to the gas, $\delta
v=|\vc{v}-\vc{u}|$. The Epstein regime is valid when the particle size is
smaller than the mean free path. The flux of impinging molecules is set in this
regime by their thermal motion and the friction time is independent of the
relative speed,
\begin{equation}
  \tau_{\rm f} = \frac{R \rho_\bullet}{c_{\rm s} \rho_{\rm g}} \, .
\end{equation}
Here $R$ is the radius of the particle, assumed to be spherical. The other
parameters are the material density $\rho_\bullet$, the gas sound speed $c_{\rm
s}$ and the gas density $\rho_{\rm g}$. Particles with sizes above $9/4$ times
the mean free path of the molecules enter the Stokes regime \citep[see][ and
references therein]{Whipple1972}, with
\begin{equation}
  \tau_{\rm f} = \frac{R \rho_\bullet}{c_{\rm s} \rho_{\rm g}} \frac{4}{9}
  \frac{R}{\lambda} \, .
\end{equation}
Here the friction time is proportional to the squared radius and independent of
gas density, since $\lambda$ is inversely proportional to the gas density. The
flow Reynolds number past the particle, ${\rm Re} = (2 R \delta v) / \nu$,
determines the further transition to drag regimes that are non-linear in the
relative speed, with the kinematic viscosity given by $\nu=(1/2) c_{\rm s}
\lambda$. At unity flow Reynolds number the drag transitions to an intermediate
regime with the friction time proportional to $(\delta v)^{-0.4}$. Above ${\rm
Re}=800$ the drag force finally becomes quadratic in the relative velocity,
with friction time
\begin{equation}
  \tau_{\rm f} = \frac{6 R \rho_{\bullet}}{(\delta v) \rho_{\rm g}} \, .
\end{equation}
Following the descriptions above, the step-wise transition from Epstein drag to
fully quadratic drag happens in the optically thin minimum mass solar nebula
\cite[MMSN,][]{Hayashi1981}, with power-law index $-1.5$ for the surface
density \citep{Weidenschilling1977b} and $-0.5$ for the temperature, at
particle sizes
\begin{eqnarray}
  R_1 = \frac{9 \lambda}{4} &=& 3.2\,{\rm cm}\,\left( \frac{r}{\rm AU}
  \right)^{2.75} \, , \label{eq:R1} \\
  R_2 = \frac{\nu}{2 (\delta v)} &\approx& 6.6\,{\rm cm}\,\left( \frac{r}{\rm
  AU} \right)^{2.5} \, , \label{eq:R2} \\
  R_3 = \frac{800 \nu}{2 (\delta v)} &\approx& 52.8\,{\rm m}\,\left(
  \frac{r}{\rm AU} \right)^{2.5} \label{eq:R3} \, .
\end{eqnarray}
Here $R_1$ denotes the Epstein-to-Stokes transition, $R_2$ the
Stokes-to-non-linear transition and $R_3$ the non-linear-to-quadratic
transition. The latter two equations are only approximate because of the
dependence of the friction time on the relative speed. Here we used the
sub-Keplerian speed $\Delta v$ as the relative speed between particle and gas,
an approximation which is only valid for large Stokes numbers (see equation
\ref{eq:deltav} below).

A natural dimensionless parameter to construct from the friction time is the
Stokes number ${\rm St}=\varOmega \tau_{\rm f}$, with $\varOmega$ denoting the
Keplerian frequency at the given orbital distance. The inverse Keplerian
frequency is the natural reference time-scale for a range of range of physical
effects in protoplanetary discs, hence the Stokes number determines (i)
turbulent collision speeds, (ii) sedimentation, (iii) radial and azimuthal
particle drift, (iv) concentration in pressure bumps and vortices and (v)
concentration by streaming instabilities. Fig.\ \ref{f:a_r_st} shows the
particle size corresponding to a range of Stokes numbers, for both the nominal
density of 1~g/cm$^3$ and extremely fluffy particles with an internal density
of $10^{-5}$~g/cm$^3$ (which could be reached when ice aggregates collide, see
section \ref{s:fluffy}).

\subsection{Radial drift}

Protoplanetary discs are slightly pressure-supported in the radial direction,
due to gradients in both mid-plane density and temperature. This leads to
sub-Keplerian motion of the gas, $v_{\rm gas} = v_{\rm K} - \Delta v$ with
$v_{\rm K}$ denoting the Keplerian speed and the sub-Keplerian velocity
difference $\Delta v$ defined as \citep{Nakagawa+etal1986}
\begin{equation}
  \Delta v \equiv \eta v_{\rm K} = - \frac{1}{2} \left( \frac{H}{r}
  \right)^2 \frac{\partial \ln P}{\partial \ln r} v_{\rm K} \, .
  \label{eq:deltav}
\end{equation}
In the MMSN the aspect ratio $H/r$ rises proportional to $r^{1/4}$ and the
logarithmic pressure gradient is $\partial \ln P/\partial \ln r=-3.25$ in the
mid-plane. This gives a sub-Keplerian speed which is constant $\Delta v=53$
m/s, independent of orbital distance. Radiative transfer models with
temperature and density dependent dust opacities yield disc aspect ratios $H/r$
with complicated dependency on $r$ and thus a sub-Keplerian motion which
depends on $r$ \citep[e.g.][]{Bell+etal1997}. Nevertheless, a sub-Keplerian
speed of $\sim$$50$ m/s can be used as the nominal value for a wide range of
protoplanetary disc models.

The drag force on the embedded particles leads to particle drift in the radial
and azimuthal directions \citep{Whipple1972,Weidenschilling1977a}
\begin{eqnarray}
  v_r    &=& - \frac{2 \Delta v}{{\rm St}+{\rm St}^{-1}} \, , \label{eq:vr} \\
  v_\phi &=& v_{\rm K} - \frac{\Delta v}{1 + {\rm St}^2} \, . \label{eq:vphi} 
\end{eqnarray}
These equations give the drift speed directly in the Epstein and Stokes
regimes. In the non-linear and quadratic drag regimes, where the Stokes number
depends on the relative speed, the equations can be solved using an iterative
method to find consistent $v_r$, $v_\phi$ and ${\rm St}$.

The azimuthal drift peaks at $v_{\phi} = v_{\rm K} - \Delta v$ for the smallest
Stokes numbers where the particles are carried passively with the sub-Keplerian
gas. The radial drift peaks at unity Stokes number where particles spiral in
towards the star at $v_r = - \Delta v$. The radial drift of smaller particles
is slowed down by friction with the gas, while particles larger than Stokes
number unity react to the perturbing gas drag by entering mildly eccentric
orbits with low radial drift.

Radial drift puts requirements on the particle growth at Stokes number around
unity (generally from ${\rm St}=0.1$ to ${\rm St}=10$) to occur within
time-scale $t_{\rm drift}$$\sim$$r/\Delta v$$\sim$100--1000 orbits
\citep{Brauer+etal2007,Brauer+etal2008a}, depending on the location in the
protoplanetary disc. However, three considerations soften the at first glance
very negative impact of the radial drift. Firstly, the ultimate fate of
drifting particles is not to fall into the star, but rather to sublimate at
evaporation fronts (or snow lines). This can lead to pile up of material
around evaporation fronts \citep{CuzziZahnle2004} and to particle growth by
condensation of water vapour onto existing ice particles
\citep{StevensonLunine1988,RosJohansen2013}. Secondly, the radial drift flow
of particles is linearly unstable to streaming instabilities
\citep{YoudinGoodman2005}, which can lead to particle concentration in dense
filaments and planetesimal formation by gravitational fragmentation of the
filaments \citep{Johansen+etal2009b,BaiStone2010b}. Thirdly, very fluffy
particles with low internal density reach unity Stokes number, where the radial
drift is highest, in the Stokes drag regime \citep{Okuzumi+etal2012}. In this
regime the Stokes number, which determines radial drift, increases as the
square of the particle size and hence growth to ``safe'' Stokes numbers with
low radial drift is much faster than for compact particles. These possibilities
are discussed more in the following sections.
\begin{figure*}[!t]
  \epsscale{0.7}
  \plotone{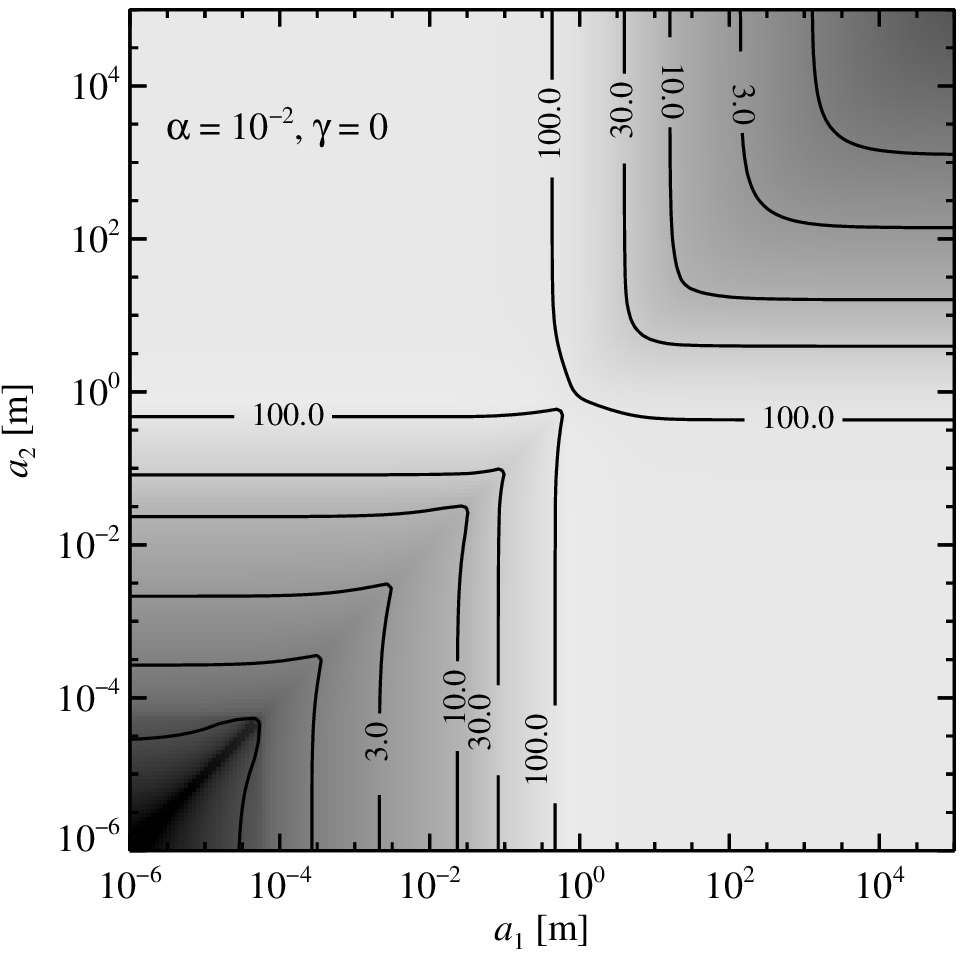}
  \plotone{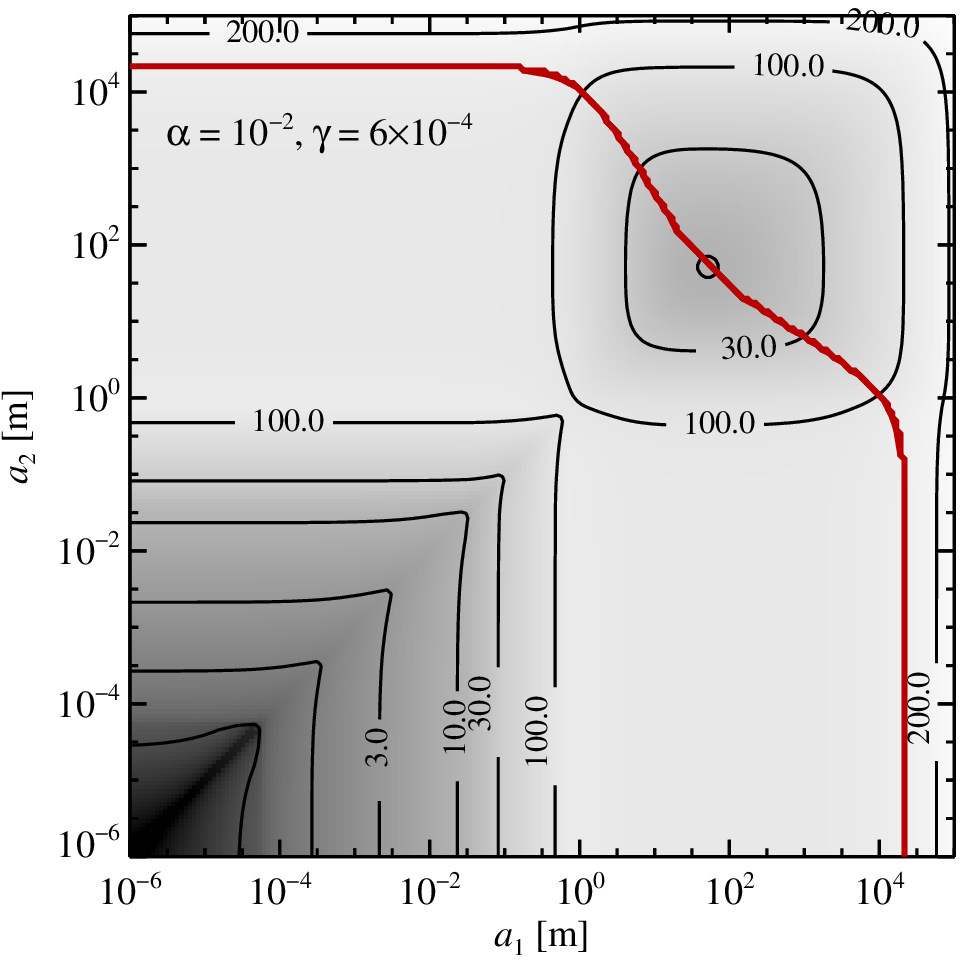} \\
  \plotone{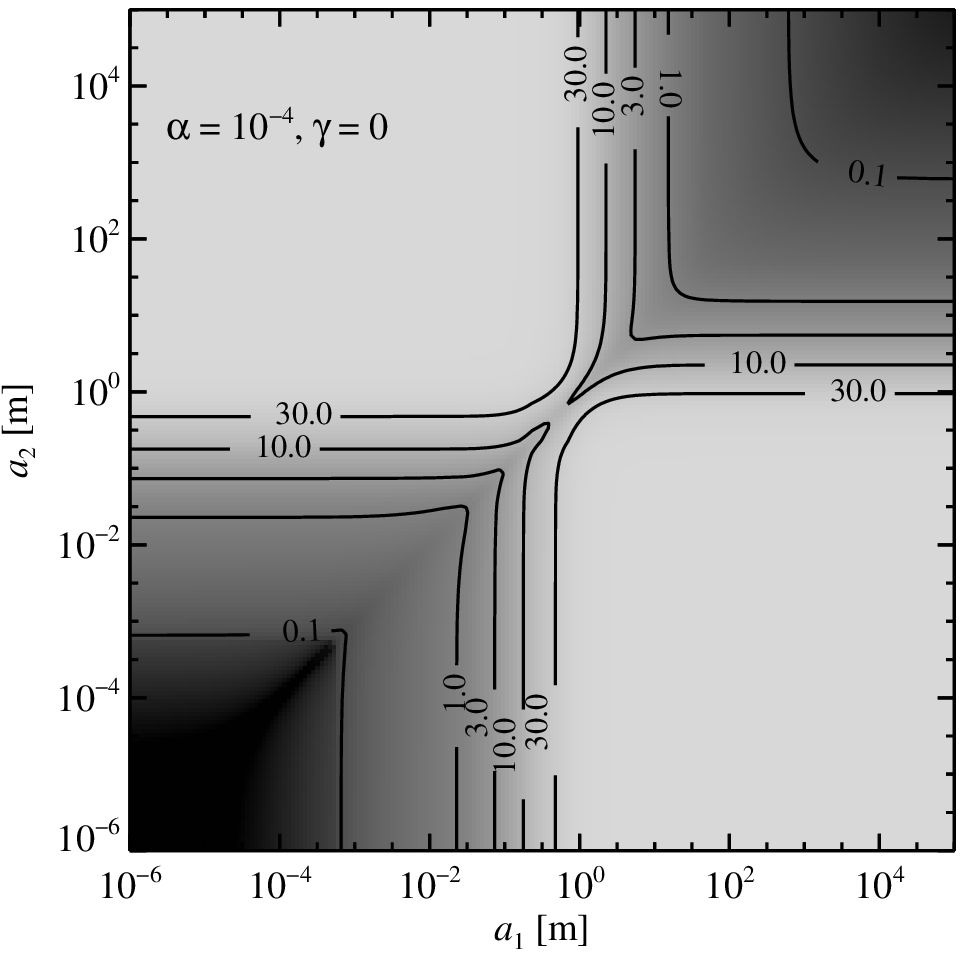}
  \plotone{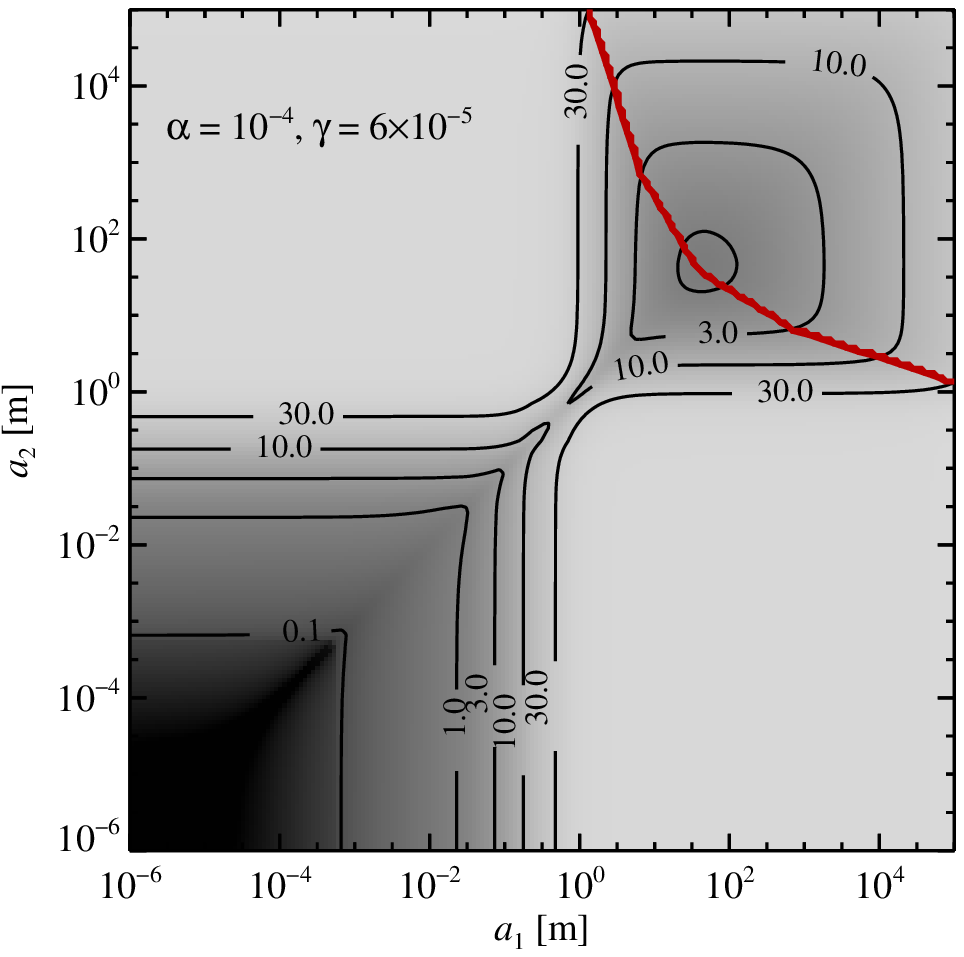}
  \caption{\small The collision speed, in meters per second, of two particles
  of size $a_1$ and $a_2$, with contributions from Brownian motion,
  differential radial and azimuthal drift, and gas turbulence. The upper panels
  show collision speeds for $\alpha=10^{-2}$ and the lower panels show
  collision speeds for $\alpha=10^{-4}$. The gravitational pull from turbulent
  gas density fluctuations is included in the right panels. The red line marks
  the transition from dominant excitation by direct drag to dominant excitation
  by turbulent density fluctuations. The ``oasis'' of low collision speeds for
  particles above 10 meters vanishes when including eccentricity pumping by
  turbulent density fluctuations.}
  \label{f:dv_a1_a2}
\end{figure*}

\subsection{Collision speeds}
\label{s:collision_speeds}

Drag from the turbulent gas excites both large-scale random motion of particles
as well as collisions (small-scale random motion). The two are distinct because
particles may have very different velocity vectors at large separations, but
these will become increasingly aligned as the particles approach each other due
to the incompressible nature of the gas flow. Particles decouple from turbulent
eddies with turn-over times shorter than the friction time, and the gradual
decoupling from the smallest eddies results in crossing particle trajectories.
This decoupling can be interpreted as singularities in the particle dynamics
\citep[so-called ``caustics'', ][]{GustavssonMehlig2011}. Caustics in turn give
rise to collisions as small particles enter regions of intense clustering
(clustering is discussed further in section \ref{s:isotropic_turbulence}). The
random motion is similar to the turbulent speed of the gas for small particles,
but particles experience decreased random motion as they grow to Stokes numbers
above unity.
\begin{figure*}
  \epsscale{1.7}
  \plotone{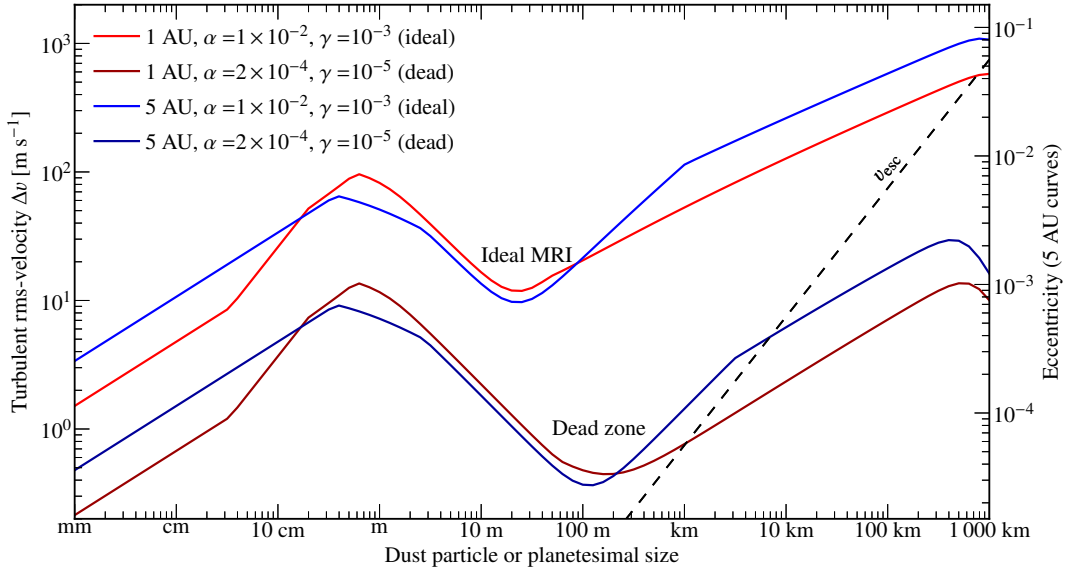}
  \caption{\small The collision speed of equal-sized particles as a function of
  their size, from dust to planetesimals, based on the stirring model of
  \cite{OrmelOkuzumi2013}. The collision speeds of small particles (below
  approximately 10 m) are excited mainly by direct gas drag, while the
  gravitational pull from turbulent gas density fluctuations dominates for
  larger particles. The transition from relative-speed-dominated to
  escape-speed-dominated can be used to define the end of the planetesimal
  formation stage (dashed line). The planetesimal sizes that must be reached
  for run-away accretion range from near 1000 km in highly turbulent discs
  ($\alpha=10^{-2}$) to around 1--10 km in disc regions with an extended dead
  zone that is stirred by sheared density waves from the active regions
  ($\alpha=2\times10^{-4}$).}
  \label{f:deltav_a}
\end{figure*}

The problem when calculating the turbulent collision speed is that at close
separations (when the particles are about to collide) they interact with the
same eddies, which causes their motions to become highly correlated. The
framework set out by \cite{Voelk+etal1980} which employs a Langevin approach,
is still widely used, and \cite{OrmelCuzzi2007} provided closed-form analytical
approximations to their results \cite[but see][ for a criticism of the
simplifications made in the V\"olk model]{PanPadoan2010}. The closed-form
expressions of \cite{OrmelCuzzi2007} require numerical solution of a single
algebraic equation for each colliding particle pair (defined by their friction
times). With knowledge of the properties of the turbulence, particularly the
turbulent rms speed and the frequency of the smallest and the largest eddies,
the collision speeds can then be calculated at all locations in the disc.

Another important contribution to turbulent collision speeds is the
gravitational pull from turbulent gas density fluctuations. The eccentricity of
a preplanetesimal increases as a random walk due to uncorrelated gravitational
kicks from the turbulent density field
\citep{Laughlin+etal2004,NelsonPapaloizou2004}. The eccentricity would grow
unbounded with time as $e \propto t^{1/2}$ in absence of dissipation. Equating
the eccentricity excitation time-scale with the time-scale for damping by tidal
interaction with the gas disc \citep[from][]{TanakaWard2004}, aerodynamic gas
drag, and inelastic collisions with other particles, \cite{Ida+etal2008}
provide parameterisations for the equilibrium eccentricity as a function of
particle mass and protoplanetary disc properties. The resulting collision
speeds dominate over the contributions from the direct drag from the turbulent
gas at sizes above approximately 10 meters.

\cite{Ida+etal2008} adopt the nomenclature of \cite{Ogihara+etal2007} for the
eccentricity evolution, where a dimensionless parameter $\gamma$ determines the
proportionality between the eccentricity and $t^{1/2}$. The parameter $\gamma$
is expected to scale with the density fluctuations $\delta \rho/\rho$ but can
be directly calibrated with turbulence simulations. The shearing box
simulations by \cite{Yang+etal2009} of turbulence caused by the
magnetorotational instability \citep{BalbusHawley1991} suggest that $\delta
\rho/\rho \propto \sqrt{\alpha}$, where $\alpha$ is the dimensionless measure
of the turbulent viscosity \citep{ShakuraSunyaev1973}. In their nominal
ideal-MHD turbulence model with $\alpha \approx 0.01$, \cite{Yang+etal2012}
find $\gamma \approx 6 \times 10^{-4}$. This leads to an approximate expression
for $\gamma$ as a function of the strength of the turbulence, $\gamma \approx
0.006 \sqrt{\alpha}$. The resulting eccentricities are in broad agreement with
the resistive magnetohydrodynamics simulations of \cite{Gressel+etal2012} where
the turbulent viscosity from Reynolds stresses fall below $\alpha=10^{-4}$ in
the mid-plane.

In Fig.\ \ref{f:dv_a1_a2} we show the collision speeds of two particles of
sizes from $\mu$m to 100 km, in a figure similar to Fig.\ 3 of the classical
{\it Protostars and Planets III} review by \cite{WeidenschillingCuzzi1993}. We
take into account the Brownian motion, the differential drift, and the gas
turbulence \citep[from the closed-form expressions of][]{OrmelCuzzi2007}. We
consider an MMSN model at $r=1\,{\rm AU}$ and $\alpha=10^{-2}$ in the upper
panels and $\alpha=10^{-4}$ in the lower panels. The gravitational pull from
turbulent density fluctuations driven by the magnetorotational instability is
included in the right panels. The collision speed approaches 100 m/s for
meter-sized boulders when $\alpha=10^{-2}$ and 30 m/s when $\alpha=10^{-4}$,
due to the combined effect of the drag from the turbulent gas and differential
drift. The collision speed of large, equal-sized particles (with ${\rm
St}$$>$$1$) would drop as
\begin{equation}
  \delta v=c_{\rm s}\sqrt{\frac{\alpha}{\rm St}}
  \label{eq:dv}
\end{equation}
in absence of turbulent density fluctuations \citep[for a discussion of this
high-St regime, see][]{OrmelCuzzi2007}. This region has been considered a safe
haven for preplanetesimals after crossing the ridges around unity Stokes number
\citep{WeidenschillingCuzzi1993}. However, the oases vanishes when including
the gravitational pull from turbulent gas density fluctuations, and instead the
collision speeds continue to rise towards larger bodies, as they are damped
less and less by gas drag (right panels of Fig.\ \ref{f:dv_a1_a2}).

The collision speed of equal-sized particles is explored further in Fig.\
\ref{f:deltav_a}. Here we show results both of a model with fully developed MRI
turbulence and a more advanced turbulent stirring model which includes the
effect of a dead zone in the mid-plane, where the ionisation degree is too low
for the gas to couple with the magnetic field
\citep{FlemingStone2003,Oishi+etal2007}, with weak stirring by density waves
travelling from the active surface layers
\citep{OkuzumiOrmel2013,OrmelOkuzumi2013}.

The collision speeds rise with size until peaking at unity Stokes number with
collision speed approximately $\sqrt{\alpha} c_{\rm s}$. The subsequent
decoupling from the gas leads to a decline by a factor ten, followed by an
increase due to the turbulent density fluctuations starting at sizes of
approximately 10 -- 100 meters. The planetesimal formation stage can be defined
as the growth until the gravitational cross section of the largest bodies is
significantly larger than their geometric cross section, a transition which
happens when the escape speed of the largest bodies approach their random
speed. In Fig.\ \ref{f:deltav_a} we indicate also the escape speed as a
function of the size of the preplanetesimal. The transition to run-away
accretion can only start at 1000 km bodies in a disc with nominal turbulence
$\alpha=10^{-2}$ \citep{Ida+etal2008}. Lower values of $\alpha$, e.g.\ in
regions of the disc where the ionisation degree is too low for the
magnetorotational instability, lead to smaller values for the planetesimal size
needed for run-away accretion, namely 1--10 km.

\subsection{Sedimentation}
\label{s:sedimentation}

While small micrometer-sized grains follow the gas density tightly, larger
particles gradually decouple from the gas flow and sediment towards the
mid-plane. An equilibrium mid-plane layer is formed when the turbulent
diffusion of the particles balances the sedimentation. The Stokes number
introduced in section \ref{s:drag_force} also controls sedimentation (because
the gravity towards the mid-plane is proportional to $\varOmega^2$). Balance
between sedimentation and turbulent diffusion \citep{Dubrulle+etal1995} yields
a mid-plane layer thickness $H_{\rm p}$, relative to the gas scale-height $H$,
of
\begin{equation}
  \frac{H_{\rm p}}{H} = \sqrt\frac{\delta}{{\rm St}+\delta} \, .
  \label{eq:Hpsed}
\end{equation}
Here $\delta$ is a measure of the diffusion coefficient $D = \delta c_{\rm s}
H$, similar to the standard definition of $\alpha$ for turbulent viscosity
\citep{ShakuraSunyaev1973}. In general $\delta \approx \alpha$ in turbulence
driven by the magnetorotational instability
\citep{JohansenKlahr2005,Turner+etal2006}, but this equality may be invalid if
accretion is driven for example by disc winds
\citep{BlandfordPayne1982,BaiStone2013}.

Equation (\ref{eq:Hpsed}) was derived assuming particles to fall towards the
mid-plane at their terminal velocity $v_z = -\tau_{\rm f} \varOmega^2 z$.
Particles with Stokes number larger than unity do not reach terminal velocity
before arriving at the mid-plane and hence undergo oscillations which are
damped by drag and excited by turbulence. Nevertheless
\cite{Carballido+etal2006} showed that equation (\ref{eq:Hpsed}) is in fact
valid for all values of the Stokes number. \cite{YoudinLithwick2007}
interpreted this as a cancellation between the increased sedimentation time of
oscillating particles and their decreased reaction to the turbulent motion of
the gas.

The particle density in the equilibrium mid-plane layer is
\begin{equation}
  \rho_{\rm p} = Z \rho_{\rm g} \frac{H}{H_{\rm p}} = Z \rho_{\rm g}
  \sqrt\frac{{\rm St}+\delta}{\delta} \, .
\end{equation}
Here $Z$ is the ratio of the particle column density to the gas column density.
In the limit of large particle sizes the mid-plane layer thickness is well
approximated by $H_{\rm p} = c_{\rm p}/\varOmega$, where $c_{\rm p}$ is the
random particle motion. This expression is obtained by associating the
collision speed of large particles in equation (\ref{eq:dv}) with their random
speed (with $\alpha=\delta$) and inserting this into equation (\ref{eq:Hpsed})
in the limit ${\rm St}\gg\delta$. The mass flux density of particles can then
be written as
\begin{equation}
  \mathcal{F} = c_{\rm p} \rho_{\rm p} = Z \rho_{\rm g} H \varOmega \, ,
\end{equation}
independent of the collision speed as well as the degree of sedimentation,
since the increased mid-plane density of larger particles is cancelled by the
decrease in collision speeds. Hence the transition from Stokes numbers above
unity to planetesimal sizes with significant gravitational cross sections is
characterised by high collision speeds and slow, ordered growth ($\dot{R}$ is
independent of size when $\mathcal{F}$ is constant) which does not benefit from
increased particle sizes nor from decreased turbulent diffusion.

Mid-plane solids-to-gas ratios above unity are reached for ${\rm St}=1$ when
$\delta<10^{-4}$. Weaker stirring allows successively smaller particles to
reach unity solids-to-gas ratio in the mid-plane. This marks an important
transition to where particles exert a significant drag on the gas and become
concentrated by streaming instabilities
\citep{YoudinGoodman2005,Johansen+etal2009b,BaiStone2010b,BaiStone2010c}. This
effect will be discussed further in the next section.

%

\section{\textbf{PARTICLE CONCENTRATION}}
\label{s:concentration}

\begin{figure*}[!t]
  \epsscale{2.0}
  \plotone{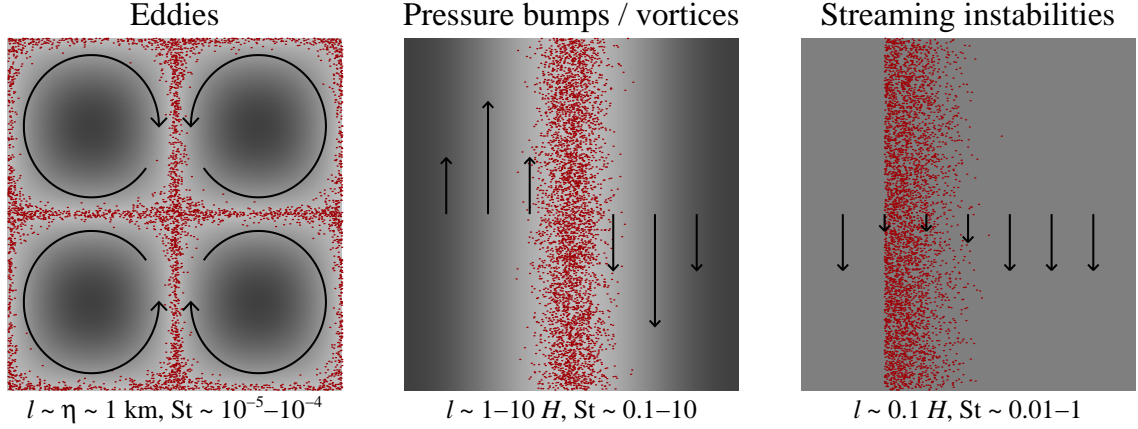}
  \caption{\small The three main ways to concentrate particles in
  protoplanetary discs. Left panel: turbulent eddies near the smallest scales
  of the turbulence, $\eta$, expel tiny particles to high-pressure regions
  between the eddies. Middle panel: the zonal flow associated with large-scale
  pressure bumps and vortices, of sizes from one scale height up to the global
  scale of the disc, trap particles of Stokes number from 0.1 to 10. Right
  panel: streaming instabilities on intermediate scales trap particles of
  Stokes number from 0.01 to 1 by accelerating the pressure-supported gas to
  near the Keplerian speed, which slows down the radial drift of particles in
  the concentration region.}
  \label{f:particle_trapping}
\end{figure*}
High local particle densities can lead to the formation of planetesimals by
gravitational instability in the sedimented mid-plane layer. Particle densities
above the Roche density
\begin{eqnarray}
  \rho_{\rm R} = \frac{9 \varOmega^2}{4 \pi G} &=& 4.3 \times
  10^{-7}\,{\rm g\,cm^{-3}}\,\left( \frac{r}{\rm AU} \right)^{-3} \nonumber \\
  &=& 315\,\rho_{\rm g} \left( \frac{r}{\rm AU} \right)^{-1/4}
\end{eqnarray}
are bound against the tidal force from the central star and can contract
towards solid densities. The scaling with gas density assumes a distribution of
gas according to the minimum mass solar nebula. Sedimentation increases the
particle density in the mid-plane and can trigger bulk gravitational
instabilities in the mid-plane layer \citep{GoldreichWard1973}. Sedimentation
is nevertheless counteracted by turbulent diffusion (see section
\ref{s:sedimentation} and equation \ref{eq:Hpsed}) and particle densities are
prevented from reaching the Roche density by global turbulence or by mid-plane
turbulence induced by the friction of the particle on the gas
\citep{Weidenschilling1980,Johansen+etal2009b,BaiStone2010b}.

High enough densities for gravitational collapse can nevertheless be reached
when particles concentrate in the gas turbulence to reach the Roche density in
local regions.

\subsection{Passive concentration}

The turbulent motion of gas in protoplanetary discs can lead to trapping of
solid particles in the flow. The size of an optimally trapped particle depends
on the length-scale and turn-over time-scale of the turbulent eddies. We
describe here particle trapping progressively from the smallest scales
unaffected by disc rotation ({\em eddies}) to the largest scales in
near-perfect geostrophic balance between the pressure gradient and Coriolis
accelerations ({\em vortices} and {\em pressure bumps}).

We consider rotating turbulent structures with length scale $\ell$, rotation
speed $v_{\rm e}$ and turn-over time $t_{\rm e} = \ell/v_{\rm e}$. These
quantities are approximate and all factors of order unity are ignored in the
following. Dust is considered to be passive particles accelerating towards the
gas velocity in the friction time $\tau_{\rm f}$, independent of the relative
velocity. Particle trapping by streaming instabilities, where the particles
play an active role in the concentration, is described in section
\ref{s:streaming}. The three main mechanisms for concentrating particles in the
turbulent gas flow in protoplanetary discs are sketched in Fig.\
\ref{f:particle_trapping}.

\subsubsection{Isotropic turbulence}
\label{s:isotropic_turbulence}

On the smallest scales of the gas flow, where the Coriolis force is negligible
over the turn-over time-scale of the eddies, the equation governing the
structure of a rotating eddy is
\begin{equation}
  \frac{{\rm d} v_r}{{\rm d} t} = - \frac{1}{\rho} \frac{\partial P}{\partial
  r} \equiv f_{\rm P} \, .
\end{equation}
Here $f_{\rm P}$ is the gas acceleration caused by the radial pressure gradient
of the eddy. We use $r$ as the radial coordinate in a frame centred on the
eddy. The pressure must rise outwards, $\partial P/\partial r>0$, to work as a
centripetal force. In such low-pressure eddies the rotation speed is set by
\begin{equation}
  f_{\rm P} = - \frac{v_{\rm e}^2}{\ell} \, .
\end{equation}
Very small particles with $\tau_{\rm f}\ll t_{\rm e}$ reach their terminal
velocity
\begin{equation}
  v_{\rm p} = -\tau_{\rm f} f_{\rm P}
\end{equation}
on a time-scale much shorter than the eddy turn-over time-scale. This gives
\begin{equation}
  v_{\rm p} = -\tau_{\rm f} f_{\rm P} = \tau_{\rm f} \frac{v_{\rm e}^2}{\ell} =
  \frac{\tau_{\rm f}}{t_{\rm e}} v_{\rm e} \, .
\end{equation}
The largest particles to reach their terminal velocity in the eddy turn-over
time-scale have $\tau_{\rm f} \sim t_{\rm e}$. This is the optimal particle
size to be expelled from small-scale eddies and cluster in regions of high
pressure between the eddies. Larger particles do not reach their terminal
velocity before the eddy structure breaks down and reforms with a new phase,
and thus their concentration is weaker.

Numerical simulations and laboratory experiments have shown that particles
coupling at the turn-over time-scale of eddies at the Kolmogorov scale of
isotropic turbulence experience the strongest concentrations
\citep{SquiresEaton1991,Fessler+etal1994}. In an astrophysics context, such
turbulent concentration of sub-mm-sized particles between small-scale eddies
has been put forward to explain the narrow size ranges of chondrules found in
primitive meteorites \citep{Cuzzi+etal2001}, as well as the formation of
asteroids by gravitational contraction of rare, extreme concentration events of
such particles \citep{Cuzzi+etal2008}. This model was nevertheless criticised
by \cite{Pan+etal2011} who found that efficiently concentrated particles have a
narrow size range and that concentration of masses sufficiently large to form
the primordial population of asteroids is hard to achieve.

\subsubsection{Turbulence in rigid rotation}

On larger scales of protoplanetary discs, gas and particle motion is dominated
by Coriolis forces and shear. We first expand our particle-trapping framework
to flows dominated by Coriolis forces and then generalise the expression to
include shear.

In a gas rotating rigidly at a frequency $\varOmega$, the equilibrium of the
eddies is now given by
\begin{equation}
  2 \varOmega v_{\rm e} - \frac{1}{\rho} \frac{\partial P}{\partial r} = - \frac{v_{\rm
  e}^2}{\ell} \, .
\end{equation}
For slowly rotating eddies with $v_{\rm e}/\ell \ll \varOmega$ we can ignore
the centripetal term and get
\begin{equation}
  v_{\rm e} = - \frac{f_{\rm P}}{2 \varOmega} \, .
\end{equation}
High pressure regions have $v_{\rm e}<0$ (clockwise rotation), while low
pressure regions have $v_{\rm e}>0$ (counter-clockwise rotation).

The terminal velocity of inertial particles can be found by solving the
equation system
\begin{eqnarray}
  \frac{{\rm d} v_x}{{\rm d} t} &=& +2 \varOmega v_y - \frac{1}{\tau_{\rm f}} v_x
  \, ,\\
  \frac{{\rm d} v_y}{{\rm d} t} &=& -2 \varOmega v_x - \frac{1}{\tau_{\rm f}}
  (v_y-v_{\rm e}) \, , 
\end{eqnarray}
for $v_x \equiv v_{\rm p}$. Here we have fixed a coordinate system in the
centre of an eddy with $x$ pointing along the radial direction and $y$ along
the rotation direction at $x=\ell$. The terminal velocity is
\begin{equation}
  v_{\rm p} = \frac{v_{\rm e}}{(2 \varOmega \tau_{\rm f})^{-1} + 2 \varOmega
  \tau_{\rm f}} \, .
  \label{eq:vp_rot}
\end{equation}
Thus high-pressure regions, with $v_{\rm e}<0$, trap particles, while
low-pressure regions, with $v_{\rm e}>0$, expel particles. The optimally
trapped particle has $2 \varOmega \tau_{\rm f}=1$. Since we are now on scales
with $t_{\rm e} \gg \varOmega^{-1}$, the optimally trapped particle has
$\tau_{\rm f} \ll t_{\rm e}$ and thus these particles have ample time to reach
their terminal velocity before the eddies turn over. An important feature of
rotating turbulence is that the optimally trapped particle has a friction time
that is {\it independent} of the eddy turn-over time-scale, and thus all eddies
trap particles of $\tau_{\rm f}\sim1/(2\varOmega)$ most efficiently.
\begin{figure*}[!t]
  \epsscale{1.9}
  \plotone{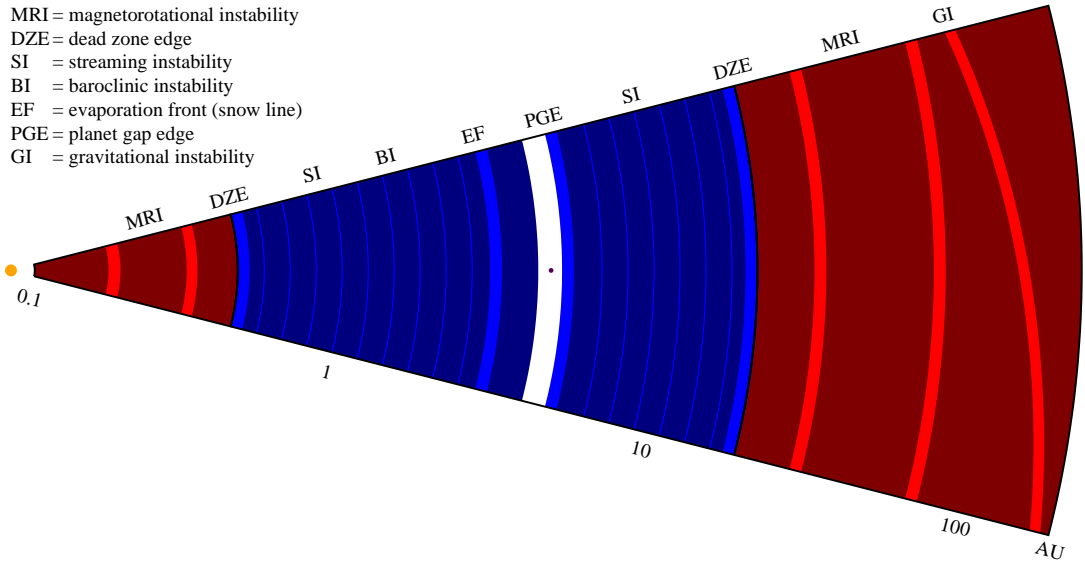}
  \caption{\small Sketch of the particle concentration regions in a wedge of a
  protoplanetary disc seen from above. Regions where the magnetorotational
  instability is expected to operate are marked with red, while the extent of
  the dead zone in a nominal protoplanetary disc model is marked with blue. The
  particle trapping mechanisms are described in the main text.}
  \label{f:particle_concentration_regions}
\end{figure*}

\subsubsection{Turbulence with rotation and shear}

Including both Keplerian shear and rotation the terminal velocity changes only
slightly compared to equation (\ref{eq:vp_rot}),
\begin{equation}
  v_{\rm p} = \frac{2 v_{\rm e}}{(\varOmega \tau_{\rm f})^{-1} + \varOmega
  \tau_{\rm f}} \, .
\end{equation}
Equilibrium structures in a rotating and shearing frame are axisymmetric
pressure bumps surrounded by super-Keplerian/sub-Keplerian zonal flows (Fig.\
\ref{f:particle_trapping}). The optimally trapped particle now has friction
time $\varOmega \tau_{\rm f}=1$. This is in stark contrast to isotropic
turbulence where each scale of the turbulence traps a small range of particle
sizes with friction times similar to the turn-over time-scale of the eddy. The
eddy speed is related to the pressure gradient through
\begin{equation}
  v_{\rm e} = -\frac{1}{2 \varOmega} \frac{1}{\rho} \frac{\partial P}{\partial r}
\end{equation}
Associating $v_{\rm e}$ with $-\Delta v$ defined in equation (\ref{eq:deltav}),
we recover the radial drift speed
\begin{equation}
  v_r = -\frac{2 \Delta v}{{\rm St}^{-1} + {\rm St}}
  \, .
\end{equation}
However, this expression now has the meaning that particles drift radially
proportional to the {\it local} value of $\Delta v$. Particles pile up where
$\Delta v$ vanishes, i.e.\ in a pressure bump.

\subsubsection{Origin of pressure bumps}

Trapping of ${\rm St}$$\sim$$1$ particles (corresponding to cm-sized pebbles to
m-sized rocks and boulders, depending on the orbital distance, see Fig.\
\ref{f:a_r_st}) in pressure bumps in protoplanetary discs has been put forward
as a possible way to cross the meter-barrier of planetesimal formation
\citep{Whipple1972,HaghighipourBoss2003}. \cite{Rice+etal2004} identified this
mechanism as the cause of particle concentration in spiral arms in simulations
of self-gravitating protoplanetary discs. Shearing box simulations of
turbulence caused by the magnetorotational instability \citep{BalbusHawley1991}
show the emergence of long-lived pressure bumps surrounded by a
super-Keplerian/sub-Keplerian zonal flow
\citep{Johansen+etal2009a,Simon+etal2012}. Similarly strong pressure bumps have
been observed in global simulations \citep{FromangNelson2005,Lyra+etal2008a}.
High-pressure anticyclonic vortices concentrate particles in the same way as
pressure bumps \citep{BargeSommeria1995}. Vortices may arise naturally through
the baroclinic instability thriving in the global entropy gradient
\citep{KlahrBodenheimer2003}, although the expression of the baroclinic
instability for realistic cooling times and in the presence of other sources of
turbulence is not yet clear
\citep{LesurPapaloizou2010,LyraKlahr2011,Raettig+etal2013}.

Pressure bumps can also be excited by a sudden jump in the turbulent viscosity
or by the tidal force of an embedded planet or star. \cite{Lyra+etal2008b}
showed that the inner and outer edges of the dead zone, where the ionisation
degree is too low for coupling the gas and the magnetic field, extending
broadly from 0.5 to 30 AU in nominal disc models \citep[][chapter by {\it
Turner et al.}]{Dzyurkevich+etal2013}, develop steep pressure gradients as the
gas piles up in the low-viscosity region. The inner edge is associated with a
pressure maximum and can directly trap particles. Across the outer edge of the
dead zone the pressure transitions from high to low, hence there is no local
maximum which can trap particles. The hydrostatic equilibrium nevertheless
breaks into large-scale particle-trapping vortices through the Rossby wave
instability \citep{Li+etal2001} in the variable-$\alpha$ simulations of
\cite{Lyra+etal2008b}. \cite{Dzyurkevich+etal2010} identified the inner
pressure bump in resistive simulations of turbulence driven by the
magnetorotational instability. The jump in particle density, and hence in
ionisation degree, at the water snow line has been proposed to cause a jump in
the surface density and act as a particle trap
\citep{KretkeLin2007,Brauer+etal2008b,Drazkowska+etal2013}. However, the snow
line pressure bump remains to be validated in magnetohydrodynamical simulations.

\cite{Lyra+etal2009a} found that the edge of the gap carved by a Jupiter-mass
planet develops a pressure bump which undergoes Rossby wave instability. The
vortices concentrate particles very efficiently. Observational evidence for a
large-scale vortex structure overdense in mm-sized pebbles was found by
\cite{vanderMarel+etal2013}. These particles have approximately Stokes number
unity at the radial distance of the vortex, in the transitional disc IRS 48.
This discovery marks the first confirmation that dust traps exist in nature.
The approximate locations of the dust-trapping mechanisms which have been
identified in the literature are shown in Fig.\
\ref{f:particle_concentration_regions}.

\subsection{Streaming instability}
\label{s:streaming}

The above considerations of passive concentration of particles in pressure
bumps ignore the back-reaction friction force exerted by the particles onto the
gas. The radial drift of particles leads to outwards motion of the gas in the
mid-plane, because of the azimuthal frictional pull of the particles on the gas.

\cite{YoudinGoodman2005} showed that the equilibrium streaming motion of gas
and particles is linearly unstable to small perturbations, a result also seen
in the simplified mid-plane layer model of \cite{GoodmanPindor2000}. The eight
dynamical equations (six for the gas and particle velocity fields and two for
the density fields) yield eight linear modes, one of which is unstable and
grows exponentially with time. The growth rate depends on both the friction
time and the particle mass-loading. Generally the growth rate increases
proportional to the friction time (up to ${\rm St}$$\sim$$1$), as the particles
enjoy increasing freedom to move relative to the gas. The dependence on the
particle mass-loading is more complicated; below a mass-loading of unity the
growth rate increases slowly (much more slowly than linearly) with
mass-loading, but after reaching unity in dust-to-gas ratio the growth rate
jumps by one or more orders of magnitude. The $e$-folding time-scale of the
unstable mode is as low as a few orbits in the regime of high mass-loading.
\begin{figure*}
  \begin{center}
    \begin{tabular}{cc}
    \vtop{\null\hbox{\includegraphics[width=0.38\linewidth]{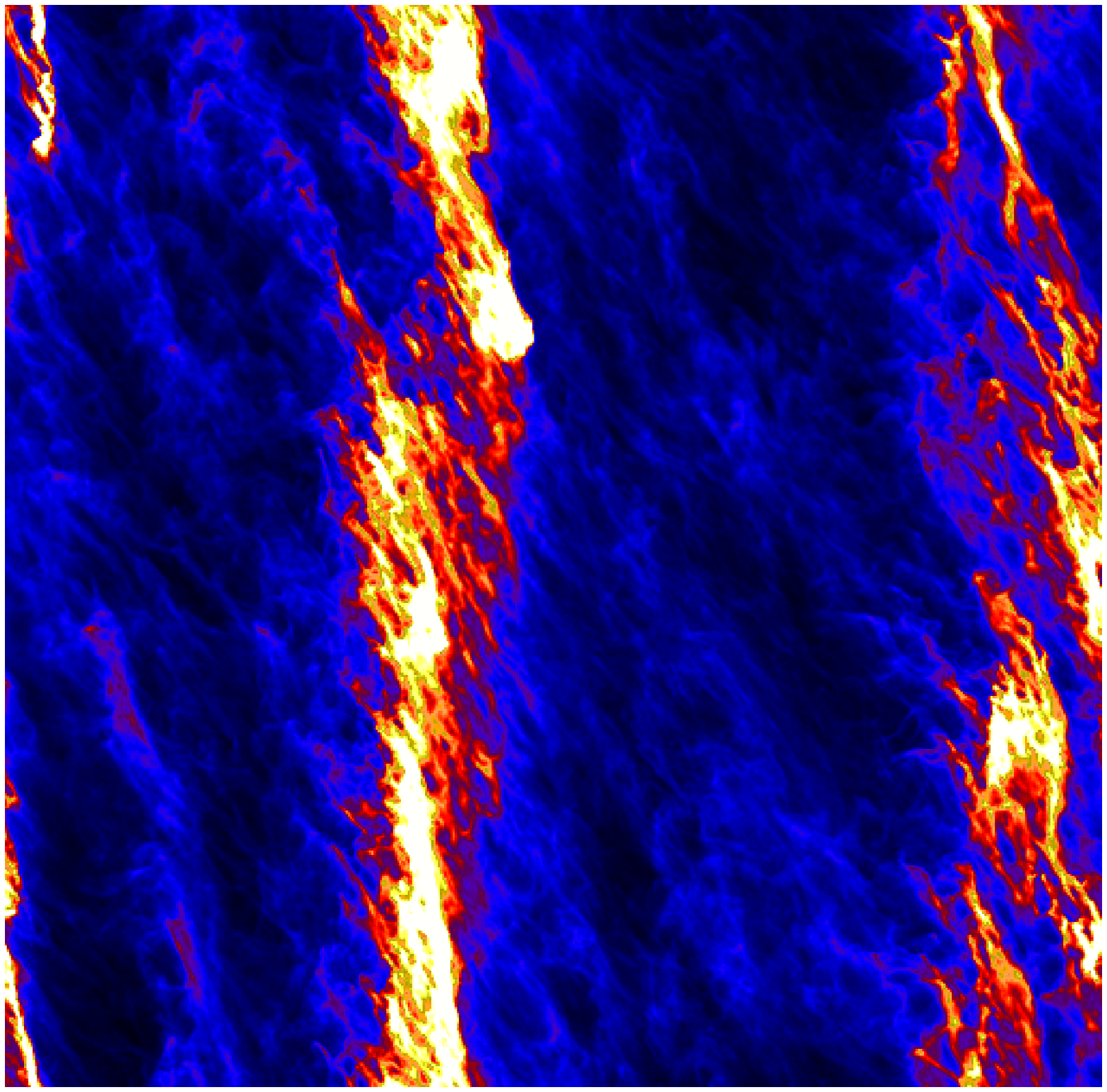}}} &
    \vtop{\null\hbox{\includegraphics[width=0.45\linewidth]{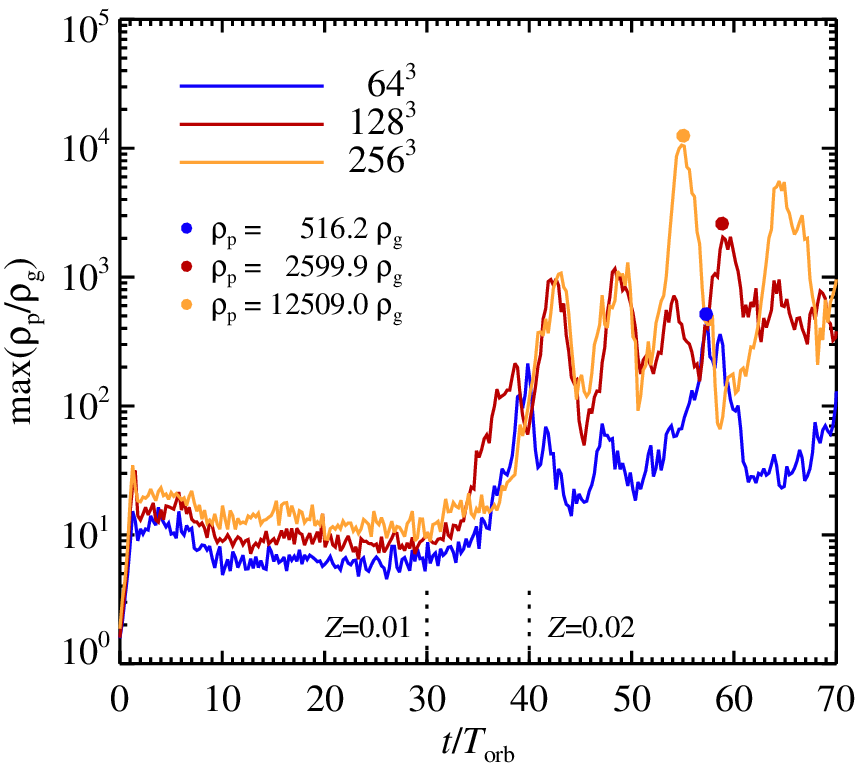}}}
    \end{tabular}
  \end{center}
  \caption{\small Particle concentration by the streaming instability is mainly
  elongated along the Keplerian flow direction, as seen in the left panel
  showing the column density of particles with ${\rm St}=0.3$ in a local frame
  where the horizontal axis represents the radial direction and the vertical
  axis the orbital direction. The substructure of the overdense region appears
  fractal with filamentary structure on many scales. The right panel shows the
  measured maximum particle density as a function of time, for three different
  grid resolutions. The metallicity is gradually increased from the initial
  $Z=0.01$ to $Z=0.02$ between 30 and 40 orbits, triggering strong particle
  concentration. Higher resolution resolves smaller embedded substructures and
  hence higher densities.}
  \label{f:rhopmax_t}
\end{figure*}

The linear mode of the streaming instability is an exact solution to the
coupled equations of motion of gas and particles, valid for very small
amplitudes. This property can be exploited to test numerical algorithm for
solving the full non-linear equations. \cite{YoudinJohansen2007} tested their
numerical algorithm for two-way drag forces against two modes of the streaming
instability, the modes having different wavenumbers and friction times. These
modes have subsequently been used in other papers \citep[e.g.][]{BaiStone2010a}
to test the robustness and convergence of numerical algorithms for the coupled
dynamics of gas and solid particles.

\subsubsection{Non-stratified simulations}

The non-linear evolution of the streaming instability can be studied either
with or without particle stratification. The case without particle
stratification is closest to the linear stability analysis. In this case the
mean particle mass-loading in the simulation domain must be specified, as well
as the friction time of the particles. The initial gas and particle velocities
are set according to drag force equilibrium \citep{Nakagawa+etal1986}, with
particle drifting in towards the star and gas drifting out.

In simulations not including the component of the stellar gravity towards the
mid-plane, i.e.\ non-stratified simulations, high particle densities are
reached mainly for particles with ${\rm St}=1$
\citep{JohansenYoudin2007,BaiStone2010a}. At a particle mass-loading of 3
(times the mean gas density), the particle density reaches almost 1000 times
the gas density. The overdense regions appear nearly axisymmetric in both local
and semi-global simulations \citep{Kowalik+etal2013}. Smaller particles with
${\rm St}=0.1$ reach only between 20 and 60 times the gas density when the
particle mass-loading is unity of higher. Little or no concentration is found
at a dust-to-gas ratio of 0.2 for ${\rm St}=0.1$ particles.
\cite{BaiStone2010a} presented convergence tests of non-stratified simulations
in 2-D and found convergence in the particle density distribution functions at
$1024^2$ grid cells.

\subsubsection{Stratified simulations}

While non-stratified simulations are excellent for testing the robustness of a
numerical algorithm and comparing to results obtained with different codes,
stratified simulations including the component of the stellar gravity towards
the mid-plane are necessary to explore the role of the streaming instability
for planetesimal formation. In the stratified case the mid-plane particle
mass-loading is no longer a parameter that can be set by hand. Rather its value
is determined self-consistently in a competition between sedimentation and
turbulent diffusion. The global metallicity $Z=\varSigma_{\rm p}/\varSigma_{\rm
g}$ is now a free parameter and the mass-loading in the mid-plane depends on
the scale height of the layer and thus on the degree of turbulent stirring.

Stratified simulations of the streaming instability display a binary behaviour
determined by the heavy element abundance. At $Z$ around the solar value or
lower, the mid-plane layer is puffed up by strong turbulence and shows little
particle concentration. \cite{BaiStone2010b} showed that the particle mid-plane
layer is stable to Kelvin-Helmholtz instabilities thriving in the vertical
shear of the gas velocity, so the mid-plane turbulence is likely a
manifestation of the streaming instability that is not associated with
particle concentration. Above solar metallicity very strong particle
concentrations occur in thin and dominantly axisymmetric filaments
\citep{Johansen+etal2009b,BaiStone2010b}. The metallicity threshold for
triggering strong clumping may be related to reaching unity particle
mass-loading in the mid-plane, necessary for particle pile-ups by the streaming
instability. The particle density can reach several thousand times the local
gas density even for relatively small particles of ${\rm St}=0.3$
\citep{Johansen+etal2012}. Measurements of the maximum particle density as a
function of time are shown in Fig.\ \ref{f:rhopmax_t}, together with a column
density plot of the overdense filaments.

Particles of ${\rm St}=0.3$ reach only modest concentration in non-stratified
simulations. The explanation for the higher concentration seen in stratified
simulations may be that the mid-plane layer is very thin, on the order of 1\%
of a gas scale height, so that slowly drifting clumps are more likely to merge
in the stratified simulations. The maximum particle concentration increases
when the resolution is increased, by a factor approximately four each time the
number of grid cells is doubled in each direction. This scaling may arise from
thin filamentary structures which resolve into thinner and thinner filaments at
higher resolution. The overall statistical properties of the turbulence
nevertheless remain unchanged as the resolution increases
\citep{Johansen+etal2012}.

The ability of the streaming instability to concentrate particles depends on
both the particle size and the local metallicity in the disc. {\it Carrera,
Johansen and Davies} (in preparation) show that particles down to ${\rm
St}=0.01$ are concentrated at a metallicity slightly above the solar value,
while smaller particles down to ${\rm St}=0.001$ require significantly
increased metallicity, e.g.\ by photoevaporation of gas or pile up of particles
from the outer disc. Whether the streaming instability can explain the presence
of mm-sized chondrules in primitive meteorites is still not known. Alternative
models based on small-scale particle concentration \citep{Cuzzi+etal2008} and
particle sedimentation \citep{YoudinShu2002,ChiangYoudin2010,Lee+etal2010} may
be necessary.
\begin{figure*}
  \epsscale{1.5}
  \includegraphics[width=\linewidth]{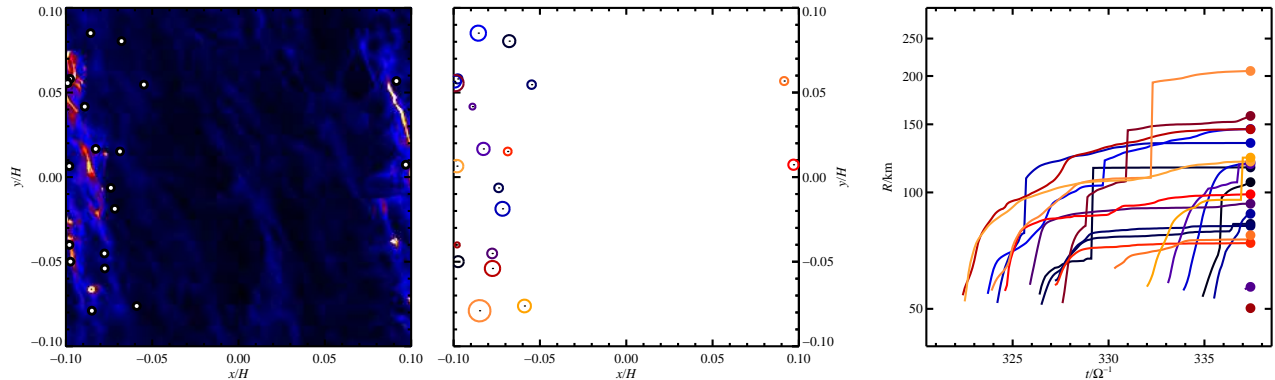}
  \caption{\small Planetesimal formation in a particle filament formed by the
  streaming instability. The left panel shows the column density of particles
  with ${\rm St}=0.3$, with dots marking the newly formed planetesimals. The
  middle panel shows the positions and Hill spheres of the planetesimals, while
  the right panel shows their contracted radii when the scale-free simulation
  is applied to the asteroid belt.}
  \label{f:planetesimals_sink}
\end{figure*}

\subsection{Gravitational collapse}
\label{s:collapse}

Planetesimals forming by gravitational contraction and collapse of the
overdense filaments are generally found to be massive, corresponding to
contracted radii between 100 km and 1000 km, depending on the adopted disc
model
\citep{Johansen+etal2007,Johansen+etal2009b,Johansen+etal2011,Johansen+etal2012,Kato+etal2012}.
Fig.\ \ref{f:planetesimals_sink} shows the results of a high-resolution
simulation of planetesimal formation through streaming instabilities, with
planetesimals from 50 to 200 km radius forming when the model is applied to
$r=3$ AU. The formation of large planetesimals was predicted already by
\cite{YoudinGoodman2005} based on the available mass in the linear modes. The
planetesimals which form increase in size with increasing disc mass
\citep{Johansen+etal2012}, with super-Ceres-sized planetesimals arising in
massive discs approaching Toomre instability in the gas
\citep{Johansen+etal2011}. Increasing the numerical resolution maintains the
size the largest planetesimals, but allows smaller-mass clumps to condense out
of the turbulent flow alongside their massive counterparts
\citep{Johansen+etal2012}. Although most of the particle mass enters
planetesimals with characteristic sizes above 100 km, it is an open question,
which can only be answered with very-high-resolution computer simulations,
whether there is a minimum size to the smallest planetesimals which can form in
particle concentration models.

Strictly speaking simulations of particle concentration and self-gravity only
find the mass spectrum of gravitationally bound clumps. Whether these clumps
will contract to form one or more planetesimals is currently not known in
details. \cite{Nesvorny+etal2010} took bound particle clumps similar to those
arising in numerical simulations and evolved in a separate $N$-body code with
collisions and merging. The rotating clumps typically collapse to a binary
planetesimal, with orbital properties in good agreement with the observed high
binary fraction in the classical cold Kuiper belt \citep{Noll+etal2008}. The
limited mass resolution which is allowed in direct $N$-body simulations
nevertheless highlights the necessity for further studies of the fate of
collapsing pebble clumps. Information from laboratory experiments on
coagulation, bouncing and fragmentation of particles is directly applicable to
self-gravitating particle clumps as well \citep{Guettler+etal2010}, and
including realistic particle interaction in collapse simulations should be a
high priority for future studies.

\section{\textbf{PARTICLE GROWTH}}
\label{s:dustgrowth}

The size distribution of solid particles in protoplanetary discs evolves due to
processes of coagulation, fragmentation, sublimation and condensation. The
particle concentration mechanisms discussed in the previous section are only
relevant for particle sizes from millimeters to meters, much larger than the
canonical sub-micron-sized dust and ice grains which enter the protoplanetary
disc. Hence growth to macroscopic sizes must happen in an environment of
relatively uniform particle densities.

Primitive meteorites show evidence of thermal processing of protoplanetary disc
solids, in the form of CAIs which likely condensed directly from the cooling
gas near the proto-Sun and chondrules which formed after rapid heating and
melting of pre-existing dust aggregates (section \ref{s:meteorites}). Thermal
processing must also have taken place near the water snow line where the
continued process of sublimation and condensation can lead to the formation of
large hail-like ice particles \citep{RosJohansen2013}; or more distantly around
the CO snow line identified observationally at 30 AU around the young star TW
Hya \citep{Qi+etal2013}.

Direct sticking is nevertheless the most general mechanism for dust growth, as
the densities in protoplanetary discs are high enough for such coagulation to
be important over the life-time of the gaseous disc. Physically, the
collisional evolution is determined by the following factors:
\begin{itemize}
  \item The spatial density of particles. This will depend on both the location
  in the disc and on the degree of particle concentration (see section
  \ref{s:concentration}).
  \item The collision cross section among particles, $\sigma_\mathrm{col}$. An
  appropriate assumption is to take spherical particles, for which
  $\sigma_{{\rm col},ij}=\pi(R_i +R_j)^2$ where $R_i$ and $R_j$ are simply the
  radii of two particles in the absence of gravitational focusing.
  \item The relative velocity among the particles, $\delta v_{ij}$. The
  relative velocity affects both the collision rate and the collision outcome.
  \item The collision outcome, which is determined by the collision energy,
  porosity, and other collision parameters like the impact parameter.
\end{itemize}
The product of $\delta v_{ij}$ and $\sigma_{ij}$, which together with the
particle density determine the collision rate, is most often referred to as the
collision kernel $K_{ij}$. 

\subsection{Numerical approaches to coagulation equation}

The evolution of the dust size distribution is most commonly described by the
Smoluchowski equation \citep{Smoluchowski1916}, which reads:
\begin{eqnarray}
  \nonumber
  \frac{dn(m)}{dt} 
  =\frac{1}{2}\int dm' K(m',m-m') n(m') n(m-m') \\
  -n(m) \int dm' K(m,m') n(m'),
  \label{eq:smol}
\end{eqnarray}
where $n(m)$ is the particle number density \textit{distribution} ({\em i.e.,}
$n(m)dm$ gives the number density of particles between mass $m$ and $m+dm$).
The terms on the right-hand side of equation (\ref{eq:smol}) simply account for
the particles of mass $m$ that are removed due to collisions with any other
particle and those gained by a collision between two particles of mass $m'$ and
$m-m'$ (the factor of 1/2 prevents double counting). In numerical applications
one most often subdivides the mass grid in bins (often logarithmically spaced)
and then counts the interactions among these bins.

However, equation (\ref{eq:smol}), an already complex integro-differential
equation, ignores many aspects of the coagulation process. For example,
collisional fragmentation is not included and the dust spatial distribution is
assumed to be homogeneous (that is, there is no local concentration). Equation
(\ref{eq:smol}) furthermore assumes that each particle pair ($m,m'$) collides
at a single velocity and ignores the effects of a velocity distribution two
particles may have. Indeed, perhaps the most important limitation of equation
(\ref{eq:smol}) is that the time-evolution of the system is assumed to be only
a function of the particles' masses. For the early coagulation phases this is
certainly inappropriate as the dust internal structure is expected to evolve
and influence the collision outcome. The collision outcomes depend very
strongly on the internal structure (porosity, fractal exponent) of dust
aggregates as well on their composition \citep[ices or silicates,
][]{Wada+etal2009,Guettler+etal2010}.

In principle the Smoluchowski coagulation equation can be extended to include
these effects \citep{Ossenkopf1993}. However, binning in additional dimensions
besides mass may render this approach impractical. To circumvent the
multi-dimensional binning \cite{Okuzumi+etal2009} outlined an extension to
Smoluchowski's equation, which treats the mean value of the additional
parameters at every mass $m$. For example, the dust at a mass bin $m$ is in
addition characterized by a filling factor $\phi$, which is allowed to change
with time. If the distribution in $\phi$ at a certain mass $m$ is expected to
be narrow, such an approach is advantageous as one may not have to deviate from
Smoluchoski's 1D framework.

A more radical approach is to use a direct-simulation Monte Carlo (MC) method,
which drops the concept of the distribution function altogether
\citep{Ormel+etal2007}. Instead the MC-method computes the collision
probability between each particle pair of the distribution (essentially the
kernel function $K_{ij}$), which depends on the properties of the particles.
Random numbers then determines which particle pair will collide and the
time-step that is involved. The outcome of the collision between these two
particles must be summarized in a set of physically-motivated recipes (the
analogy with equation \ref{eq:smol} is simply $m_1+m_2\rightarrow m$). A new
set of collision rates is computed, after which the procedure repeats itself.

Since there are many more dust grains in a disc than any computer can handle,
the computational particles should be chosen such to accurately
\textit{represent} the physical size distribution. A natural choice is to
sample the mass of the distribution \citep{ZsomDullemond2008}. This method
also allows fragmentation to be naturally incorporated. The drawback however is
that the tails of the size distribution are not well resolved. The limited
dynamic range is one of the key drawbacks of MC-methods.
\cite{OrmelSpaans2008} have described a solution, but at the expense of a more
complex algorithm.

\subsection{Laboratory experiments}

Numerical solutions to the coagulation equation rely crucially on input from
laboratory experiments on the outcome of collisions between dust particles.
Collision experiments in the laboratory as well as under microgravity
conditions over the past 20 years have proven invaluable for the modeling of
the dust evolution in protoplanetary discs. Here we briefly review the
state-of-the-art of these experiments (more details are given in the chapter by
{\it Testi et al.}). A detailed physical model for the collision behavior of
silicate aggregates of all masses, mass ratios and porosities can be found in
\cite{Guettler+etal2010} and recent modifications are published by
\cite{Kothe+etal2013}.

The three basic types of collisional interactions between dust aggregates are:
\begin{enumerate}
  \item Direct collisional sticking: when two dust aggregates collide gently
  enough, contact forces (e.g., van der Waals forces) are sufficiently strong
  to bind the aggregates together. Thus, a more massive dust aggregate has
  formed. Laboratory experiments summarized by \cite{Kothe+etal2013} suggest
  that there is a mass-velocity threshold $v \propto m^{-3/4}$ for spherical
  medium-porosity dust aggregates such that only dust aggregates less massive
  and slower than this threshold can stick. However, hierarchical dust
  aggregates still stick at higher velocities/masses than given by the
  threshold for homogeneous dust aggregates \citep{Kothe+etal2013}.
  \item Fragmentation: when the collision energy is sufficiently high,
  similar-sized dust aggregates break up so that the maximum remaining mass
  decreases under the initial masses of the aggregates and a power-law type
  tail of smaller-mass fragments is produced. The typical fragmentation
  velocity for dust aggregates consisting of $\mu m$-sized silicate monomer
  grains is $1~\rm m~s^{-1}$.
  \item Bouncing: above the sticking threshold and below the fragmentation
  limit, dust aggregates bounce off one another. Although the conditions under
  which bouncing occurs are still under debate (see sections
  \ref{s:coagulation-fragmentation} and \ref{s:fluffy}), some aspects of
  bouncing are clear. Bouncing does not directly lead to further mass gain so
  that the growth of dust aggregates is stopped (``bouncing barrier'').
  Furthermore, bouncing leads to a steady compression of the dust aggregates so
  that their porosity decreases to values of typically 60\%
  \citep{Weidling+etal2009}.
\end{enumerate}
Although the details of the three regimes are complex -- regime boundaries are
not sharp, and other more subtle processes exist -- this simplified physical
picture shows that in protoplanetary discs, in which the mean collision
velocity increases with increasing dust-aggregate size (see Fig.\
\ref{f:dv_a1_a2}), a {\em maximum aggregate mass} exists. Detailed numerical
simulations using the dust-aggregate collision model by
\cite{Guettler+etal2010} find such a maximum mass \citep{Zsom+etal2010,
Windmark+etal2012a}. For a minimum mass solar nebula model at 1 AU, the maximum
dust-aggregate size is in the range of millimeters to centimeters
\citep[e.g.][]{Zsom+etal2010}.

Even if the bouncing barrier can be overcome, another perhaps even more
formidable obstacle will present itself at the size scale where collision
velocities reach $\sim$1 m/s (Fig.\ \ref{f:dv_a1_a2}). At this size collisions
among two silicate, similar-size, dust particles are seen to fragment, rather
than accrete. Consequently, growth will stall and the size distribution will,
like with the bouncing case, settle into a steady-state
\citep{Birnstiel+etal2010,Birnstiel+etal2011,Birnstiel+etal2012}. Highly porous
icy aggregates may nevertheless still stick at high collision speeds (see
section \ref{s:fluffy}).

\section{\textbf{GROWTH BY MASS TRANSFER}}
\label{s:coagulation-fragmentation}

Over the last two decades a large effort has been invested to study
experimentally the collisions among dust particles and to assess the role of
collisions in planetesimal formation. As was seen in section \ref{s:dustgrowth}
a plethora of outcomes -- sticking, compaction, bouncing, fragmentation -- is
possible depending mainly on the collision velocity and the particle size
ratio. Using the most updated laboratory knowledge of collision outcomes,
\cite{Zsom+etal2010,Zsom+etal2011} solved the coagulation equation with a Monte
Carlo method and found that in the ice-free, inner disc regions the dust size
distribution settles into a state dominated by mm-size particles. Further
growth is impeded because collisions among two such mm-size particles will
mainly lead to bouncing and compactification.
\begin{figure}[!t]
  \epsscale{1.0}
  \plotone{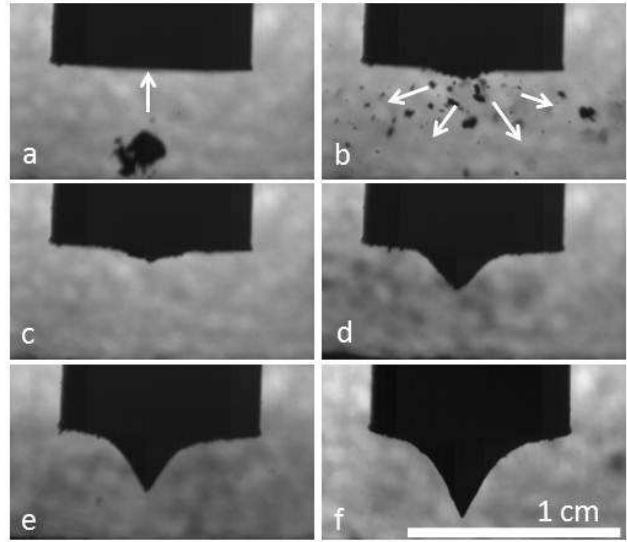}
  \caption{\small Experimental example of mass transfer in fragmenting
  collisions. All experiments were performed in vacuum. (a) A mm-sized fluffy
  dust aggregate is ballistically approaching the cm-sized dusty target at a
  velocity of 4.2 m/s. Projectile and target consist of monodisperse SiO$_2$
  spheres of 1.5 $\mu$m diameter. (b) Shortly after impact, most of the
  projectile's mass flies off the target in form of small fragments (as
  indicated by the white arrows); part of the projectile sticks to the target.
  (c) - (e) The same target after 3 (c), 24 (d), 74 (e) and 196 (f) consecutive
  impacts on the same spot. Image credit: Stefan Kothe, TU Braunschweig.}
  \label{f:mass_transfer}
\end{figure}

The real problem behind this ``bouncing barrier'' may, counter-intuitively, not
be the bouncing but rather the absence of erosion and fragmentation events.
Based upon laboratory experiments
\citep{Wurm+etal2005,TeiserWurm2009a,TeiserWurm2009b,Kothe+etal2010,
Teiser+etal2011,Meisner+etal2013}, it has become clear that above the
fragmentation limit, the collision between two dust aggregates can lead to a
mass gain of the larger (target) aggregate if the smaller (projectile)
aggregate is below a certain threshold. For impact velocities in the right
range and relatively small projectile aggregates, up to 50\% of the mass of the
projectile can be firmly transferred to the target. This process has been shown
to continue to work after multiple collisions with the target and under a wide
range of impact angles (see Fig.\ \ref{f:mass_transfer}).

Growth by mass transfer requires a high fraction of the dust mass present in
small dust aggregates which can be swept up by the larger particles. The number
of small dust aggregates can be maintained by catastrophic collisions between
larger bodies. A simple two-component model for such coagulation-fragmentation
growth was developed by \cite{Johansen+etal2008}. They divided the size
distribution into small particles and large particles and assumed that (1) a
small particle will stick to a large particle, (2) a small particle will bounce
off another small particle, and (3) a large particle will shatter another large
particle. Small particles collide with large particles at speed $v_{12}$, while
large particles collide with each other at speed $v_{22}$. Under these
conditions an equilibrium in the surface density ratio of large and small
particles, $\varSigma_1/\varSigma_2$, can be reached, with equal mass flux from
large to small particles by fragmentation as from small to large particles by
mass transfer in collisions.

In the equilibrium state where the mass flux from large to small particles
balances the mass flux from small to large particles, the growth rate of the
large particles is
\begin{eqnarray}
  \dot{R} &=& \frac{\varSigma_{\rm p}/(\sqrt{2\pi}H_1)}{\rho_\bullet}
  \times \nonumber \\ & &
  \frac{v_{22}}{[2/(1+H_1^2/H_2^2)]^{1/2}+4v_{22}/v_{12}} \, .
\end{eqnarray}
Here $H_1$ and $H_2$ are the scale heights of the small particles and the large
particles, respectively, and $\varSigma_{\rm p}=\varSigma_1+\varSigma_2$ is the
total column density of the dust particles. The complexity of the equation
arises from the assumption that the small grains are continuously created in
collisions between the large particles. Setting $H_1=H$ and $H_2 \ll H_1$ the
equations are valid for a very general sweep-up problem where a few large
particles sweep up a static population of small particles, resulting in growth
of the large particles at rate
\begin{eqnarray}
  \dot{R} &=& \frac{\epsilon Z \rho_{\rm g}}{4 \rho_\bullet} v_{12} = 2.7 \,
  {\rm mm\,yr^{-1}} \, \left( \frac{\epsilon}{0.5} \right) \left( \frac{r}{\rm
  AU} \right)^{-2.75} \times \nonumber \\ & & \hspace{-0.5cm} \left(
  \frac{Z}{0.01} \right) \left( \frac{\rho_\bullet}{1\,{\rm g\,cm^{-3}}}
  \right)^{-1} \left( \frac{v_{12}}{50\,{\rm m\,s^{-1}}} \right) \, .
  \label{eq:R2dotsimple}
\end{eqnarray}
We introduced here a sticking coefficient $\epsilon$ which measures either the
sticking probability or the fraction of mass which is transferred from the
projectile to the target. The growth rate is constant and depends only on the
particle size through $v_{12}$. If the small particles are smaller than Stokes
number unity and the large particles are larger, then we can assume that the
flux of small particles is carried with the sub-Keplerian wind, $v_{12} \approx
\Delta v$, and solve equation (\ref{eq:R2dotsimple}) for the time-scale to
cross the radial drift barrier. The end of the radial drift barrier can be set
at reaching roughly ${\rm St}=10$. This Stokes number corresponds in the
Epstein regime to the particle size
\begin{equation}
  R_{10}^{\rm (Ep)} = \frac{10 H \rho_{\rm g}}{\rho_\bullet} \, .
\end{equation}
In the Stokes regime the particle size is
\begin{equation}
  R_{10}^{\rm (St)} = \sqrt{\frac{90 H \rho_{\rm g} \lambda}{4 \rho_\bullet}}
  \, .
\end{equation}
Fig.~\ref{f:a_r_st} shows that compact ice particles grow to Stokes number 10
in the Epstein regime outside of $r=5$ AU and in the Stokes or non-linear
regimes inside of this radius. Fluffy ice particles with a very low internal
density of $\rho_\bullet=10^{-5}$ g cm$^{-3}$ grow to Stokes number 10 in the
non-linear or quadratic regimes in the entire extent of the disc. In the
Epstein regime the time to grow to Stokes number 10,
$\tau_{10}=R_{10}/\dot{R}$, is independent of both material density and gas
density,
\begin{equation}
  \tau_{10}^{\rm (Ep)} = \frac{40 H}{\epsilon Z v_{12}} \approx 25000\,{\rm
  yr}\, \left( \frac{\epsilon}{0.5} \right)^{-1} \left(\frac{r}{\rm
  AU}\right)^{5/4} \, .
  \label{eq:eps_t10}
\end{equation}
The growth-time is much shorter in the Stokes regime where the Stokes number is
proportional to the squared particle radius, giving
\begin{eqnarray}
  \tau_{10}^{\rm (St)} &=& \frac{\sqrt{360 H \rho_\bullet (m/\sigma)}}{
  \epsilon Z \rho_{\rm g} v_{12}} \approx 500\,{\rm yr} \times \nonumber \\
  & &\left( \frac{\epsilon}{0.5}\right)^{-1} \left( \frac{\rho_\bullet}{1\,{\rm
  g\,cm^{-3}}} \right)^{1/2} \left( \frac{r}{\rm AU} \right)^{27/8} \, ,
  \label{eq:t10st}
\end{eqnarray}
with a direct dependence on both material density and gas density. Here $m$ and
$\sigma$ are the mass and collisional cross section of a hydrogen molecule.
Regarding the sticking coefficient, \cite{Wurm+etal2005} find mass transfer
efficiencies of $\epsilon\approx0.5$ for collision speeds up to $v_{12}=25$
m/s, hence we have used $\epsilon=0.5$ as a reference value in the above
equations. It is clearly advantageous to have a low internal density and cross
the radial drift barrier in the Stokes regime \citep{Okuzumi+etal2012}. The
non-linear and quadratic regimes are similarly beneficial for the growth, with
the time-scale for crossing the meter barrier in the quadratic drag force
regime approximately a factor $6 c_{\rm s}/\Delta v$ faster than in the Epstein
regime (equation \ref{eq:eps_t10}).

The two-component model presented in \cite{Johansen+etal2008} is analytically
solvable, but very simplified in the collision physics.
\cite{Windmark+etal2012a} presented solutions to the full coagulation equation
including a velocity-dependent mass transfer rate. They showed that
artificially injected seeds of centimeter sizes can grow by sweeping up smaller
mm-sized particles which are stuck at the bouncing barrier
\citep{Zsom+etal2010}, achieving at 3 AU growth to 3 meters in 100,000 years
and to 100 meters in 1 million years. The time-scale to grow across the radial
drift barrier is broadly in agreement with the simplified model presented above
in equation (\ref{eq:t10st}). This time is much longer than the radial drift
time-scale which may be as short as 100--1,000 years. It is therefore necessary
to invoke pressure bumps to protect the particles from radial drift while they
grow slowly by mass transfer. However, in that case the azimuthal drift of
small particles vanishes and a relatively high turbulent viscosity of
$\alpha=10^{-2}$ must be used to regain the flux onto the large particles. The
pressure bumps must additionally be very long-lived, on the order of the
life-time of the protoplanetary disc.

While the discovery of a pathway for growth by mass transfer at high-speed
collisions is a major experimental breakthrough, it seems that this effect can
not in itself lead to widespread planetesimal formation, unless perhaps in the
innermost part of the protoplanetary disc where dynamical time-scales are short
(equation \ref{eq:t10st}). The planetesimal sizes which are reached within a
million years are also quite small, on the order of 100 meters. These sizes are
too small to undergo gravitational focusing even in weakly turbulent discs (see
Fig.\ \ref{f:deltav_a}). An additional concern is the erosion of the
preplanetesimal by tiny dust grains carried with the sub-Keplerian wind
\citep{SchraeplerBlum2011,Seizinger+etal2013}.

Particles stuck at the bouncing barrier are at the lower end of the sizes that
can undergo particle concentration and gravitational collapse. However, if a
subset of ``lucky'' particles experience only low-speed, sticking collisions
and manage to grow past the bouncing barrier, this can eventually lead to the
crossing of the bouncing barrier by a very small fraction of the particles
\citep{Windmark+etal2012b,Garaud+etal2013}. This break-through nevertheless
still requires a pressure bump to stop the radial drift.

\section{\textbf{FLUFFY GROWTH}}
\label{s:fluffy}

The last years have seen major improvements in $N$-body molecular-dynamics
simulations of dust aggregate collisions by a number of groups
\citep[e.g.][]{Wada+etal2008,Wada+etal2009,Wada+etal2011,PaszunDominik2008,PaszunDominik2009,SeizingerKley2013}.
These $N$-body simulations show that fluffy aggregates have the potential to
overcome barriers in dust growth. In $N$-body simulations of dust aggregates,
all surface interactions between monomers in contact in the aggregates are
calculated, by using a particle interaction model.

$N$-body simulations of dust aggregates have some merits compared to laboratory
experiments. Precise information such as the channels through which the
collisional energy is dissipated can be readily obtained and subsequent data
analysis is straightforward. Another advantage of $N$-body simulations is that
one can study highly fluffy aggregates, which are otherwise crushed under the
Earth's gravity. On the other hand, an accurate interaction model of
constituent particles is required in $N$-body simulations of dust aggregates.
In the interaction model used in most $N$-body simulations, the constituent
particles are considered as adhesive elastic spheres. The adhesion force
between them is described by the JKR theory \citep{Johnson+etal1971}. As for
tangential resistive forces against sliding, rolling, or twisting motions,
Dominik and Tielens's model is used
\citep{DominikTielens1995,DominikTielens1996,DominikTielens1997,Wada+etal2007}.
In order to reproduce the results of laboratory experiments better, the
interaction model must be calibrated against laboratory experiments
\citep{PaszunDominik2008,Seizinger+etal2012,Tanaka+etal2012}.

In protoplanetary discs, dust aggregates are expected to have fluffy structures
with low bulk densities if impact compactification is negligible
\citep[e.g.][]{Okuzumi+etal2009,Okuzumi+etal2012,Zsom+etal2010,Zsom+etal2011}.
Especially in the early stage of dust growth, low-speed impacts result in
hit-and-stick of dust aggregates, with little or no compression. This growth
mode makes fluffy aggregates with a low fractal dimension of $\sim$$2$. Thus
it is necessary to examine the outcome of collisions between fluffy dust
aggregates and the resulting compression. The results of $N$-body simulations
can then be used in numerical models of the coagulation equation including the
evolution of the dust porosity.

\subsection{Critical impact velocity for dust growth}

\cite{DominikTielens1997} first carried out $N$-body simulation of aggregate
collisions, using the particle interaction model they constructed. They also
derived a simple recipe for outcomes of aggregate collisions from their
numerical results, though they only examined head-on collisions of
two-dimensional small aggregates containing as few as 40 particles. The
collision outcomes are classified into hit-and-stick, sticking with
compression, and catastrophic disruption, depending on the impact energy.
Bouncing of two aggregates was not observed in their simulations.
\begin{figure}[!t]
  \includegraphics[width=\linewidth]{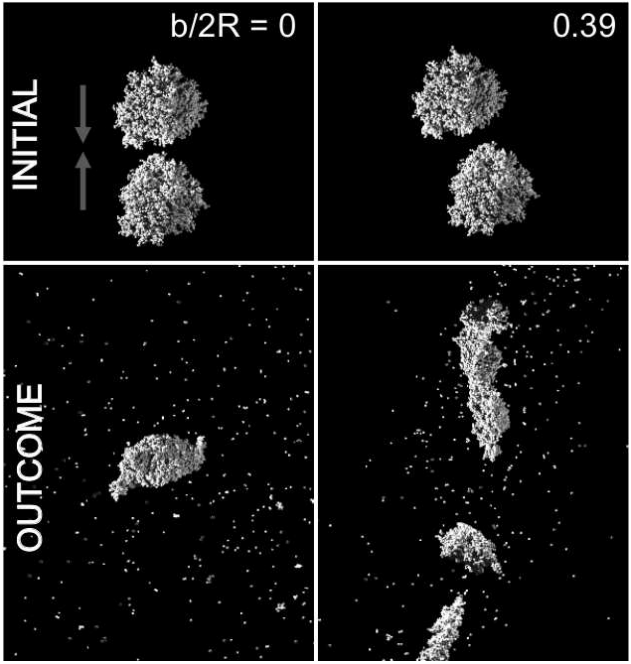}
  \caption{\small Examples of collision outcomes of icy PCA clusters consisting
  of 8000 particles for two values of the impact parameter $b$ (lower panels).
  The upper panels represent initial aggregates. The collision velocity of 70
  m/s in both cases and $R$ is the radius of the initial aggregates. The
  head-on collision (left panel) results in sticking with minor fragmentation
  while aggregates tend to pass by each other in off-set collisions (right
    panels). (This figure is reproduced based on Fig.\ 2 of Wada et al. 2009 by
    permission of the AAS.)}
  \label{f:fig_tanaka}
\end{figure}

According to the DT \citep{DominikTielens1997} recipe, growth is possible at
collisions with $E\sub{imp} < A n_k E\sub{break}$, where $A$$\sim$$10$ and
$n_k$ is the total number of contacts in two colliding aggregates. For
relatively fluffy aggregates, $n_k$ is approximately equal to the total number
of constituent particles in the two aggregates, $N$. The energy for breaking
one contact between two identical particles, $E\sub{break}$, is given by
\citep[e.g.][]{Chokshi+etal1993,Wada+etal2007}
\begin{equation}
  E\sub{break} = 23 [ \gamma^5 r^4 (1-\nu^2)^2 /\mathcal{E}^2]^{1/3},
  \label{eq1-tanaka}
\end{equation}
where $r$, $\gamma$, $\mathcal{E}$, and $\nu$ are the radius, the surface
energy, Young's modulus, and Poisson's ratio of constituent particles,
respectively. For icy particles the breaking energy is obtained as $5.2 \times
10^{-10} (r/0.1\,{\rm \mu m})^{4/3}$ erg while the value for silicate particles
is $1.4 \times 10^{-11} (r/0.1{\rm \mu m})^{4/3}$ erg, so icy aggregates are
much more sticky than silicate aggregates

Using the impact velocity, $v\sub{imp}$, the impact energy is expressed as
$(Nm/4)v\sub{imp}^2/2$, where $m$ is the mass of a constituent particle and
$Nm/4$ is the reduced mass of two equal-sized aggregates. Substituting this
expression into the above energy criterion for dust growth and noting that
$n_k\simeq N$, we obtain the velocity criterion as
\begin{equation}
  v\sub{imp} < \sqrt{8AE\sub{break}/m}.
  \label{eq2-tanaka}
\end{equation}
Here $A$ is a dimensionless parameter that needs to be calibrated with
experiments. Note that this velocity criterion is dependent only on properties
of constituent particles and, thus, independent of the mass of the aggregates.

To evaluate the critical impact velocity for growth more accurately,
\cite{Wada+etal2009} performed $N$-body simulations of large aggregates made of
up to $\sim$$10^4$ sub-micron icy particles, including off-set collisions (see
Fig.\ \ref{f:fig_tanaka}). They consider two kinds of aggregate structure, the
so-called CCA and PCA clusters. The CCA (Cluster-Cluster Agglomeration)
clusters have an open structure with a fractal dimension of 2 while the PCA
(Particle-Cluster Agglomeration) clusters have a fractal dimension of 3 and a
volume filling factor of 0.15. The PCA clusters are rather compact, compared to
the CCA. Since dust aggregates are expected to be much more compact than CCA
clusters due to compression (see section \ref{s:compression}), the growth and
disruption process of PCA clusters is of particular importance for elucidating
the planetesimal formation.

For PCA clusters (composed of 0.1$\mu$m-sized icy particles), the critical
impact velocity is obtained as 60 m/s from their simulations, independent of
the aggregate mass within the mass range examined in the simulations of
\cite{Wada+etal2009}. This indicates that icy dust aggregates can circumvent
the fragmentation barrier and grow towards planetesimal sizes via collisional
sticking. Note that the critical velocity actually increases with the aggregate
mass when only head-on collisions are considered, as seen in Fig.\
\ref{f:fig_tanaka}. For an accurate evaluation of the critical velocity for
growth, off-set collisions should be considered as well, as in
\cite{Wada+etal2009}.

The above result also fixes the corresponding constant $A$ in the critical
energy at 30. By using $E_{\rm break}$ of silicate particles in equation
(\ref{eq2-tanaka}) with $A=30$, the critical velocity for growth of silicate
aggregates is obtained as $v_{\rm imp}=1.3\,(r/0.6\mu\mbox{m})^{-5/6}$ m/s. It
agrees well with the laboratory experiments and numerical simulations for small
silicate aggregates
\citep[e.g.][]{BlumWurm2000,Guettler+etal2010,PaszunDominik2009,SeizingerKley2013}.

In the case of CCA clusters, \cite{Wada+etal2009} found that the constant $A$
is the same as in the DT recipe ($\simeq$10). Considering much smaller volume
filling factors of CCA clusters than PCA, it indicates that the critical
velocity is only weakly dependent on the volume filling factor.

\subsection{Compression of dust aggregates}
\label{s:compression}

The DT recipe does not describe the amount of changes in the porosity (or the
volume filling factor) at aggregate collisions. The first attempt to model
porosity changes was done by \cite{Ormel+etal2007}, using simple prescriptions
for the collision outcome of porosity.

\cite{Wada+etal2008} examined compression at head-on collisions of two equal
sized CCA clusters, using a high number of $N$-body simulations. Compression
(or restructuring) of an aggregate occurs through rolling motions between
constituent particles. Thus it is governed by the rolling energy $E\sub{roll}$
(i.e., the energy for rolling of a particle over a quarter of the circumference
of another particle in contact). At low-energy impacts with $E\sub{imp}
\lesssim E\sub{roll}$, aggregates just stick to each other, as indicated by the
DT recipe. For higher-energy impacts, the resultant aggregates are compressed,
depending on the impact energy. The radius of resulting aggregates, $R$, is
fitted well with the power-law function
\begin{equation}
  R \simeq 0.8 [E\sub{imp}/(NE\sub{roll})]^{-0.1}N^{1/2.5} r \, ,
  \label{eq3-tanaka}
\end{equation}
where $r$ is the radius of a monomer and $N$ the number of constituent
particles. Analysing the structure of the compressed aggregates, it can be
shown that the compressed aggregates have a fractal dimension of 2.5. This
fractal dimension is consistent with equation~(\ref{eq3-tanaka}) since
equation~(\ref{eq3-tanaka}) gives the relation of $N \propto R^{2.5}$ for
maximally compressed aggregates (with $E\sub{imp}$$\sim$$NE\sub{roll}$). The
compression observed in the $N$-body simulation is less much extreme than in
the simple porosity model used by \cite{Ormel+etal2007} because of the low
fractal dimension of 2.5.

While \cite{Wada+etal2008} examined the compression in a single collision, dust
aggregates will be gradually compressed by successive collisions in realistic
systems. In order to examine such a gradual compression process during growth,
\cite{Suyama+etal2008} performed $N$-body simulations of sequential collisions
of aggregates. Even after multiple collisions, the compressed aggregates
maintain a fractal dimension of 2.5. \cite{Suyama+etal2012} further extend
their porosity model to unequal-mass collisions.

\cite{Okuzumi+etal2012} applied the porosity evolution model of
\cite{Suyama+etal2012} to dust growth outside of the snow line in a
protoplanetary disc. As a first step, collisional fragmentation is neglected
because a relatively high impact velocity ($\sim$60 m/s) is required for
significant disruption of icy dust aggregates. They found that dust particles
evolve into highly porous aggregates (with bulk densities much less than 0.1
g/cm$^3$) even if collisional compression is taken into account. This is due to
the ineffective compression at aggregate collisions in the porosity model of
\cite{Suyama+etal2012}. Another important aspect of fluffy aggregates is that
they cross the radial drift barrier in the Stokes or non-linear drag force
regimes (see Fig.\ \ref{f:a_r_st} and discussion in section
\ref{s:coagulation-fragmentation}) where it is easier to grow to sizes where
radial drift is unimportant. This mechanism accelerates dust growth at the
radial drift barrier effectively and enables fluffy aggregates to overcome this
barrier inside 10 AU \citep{Okuzumi+etal2012,Kataoka+etal2013}.

\subsection{Bouncing condition in $N$-body simulations}

Until 2011, bouncing events were not reported in N-body simulations of
aggregate collisions, while bouncing events have been frequently observed in
laboratory experiments \citep[e.g.][]{BlumMuench1993,Langkowski+etal2008,
Weidling+etal2009,Weidling+etal2012}. \cite{Wada+etal2011} found bouncing
events in their N-body simulations of both icy and silicate cases for rather
compact aggregates with filling factor $\phi \ge 0.35$. Such compact aggregates
have a relatively large coordination number (i.e., the mean number of particles
in contact with a particle), which inhibits the energy dissipation through the
rolling deformation and helps bouncing. \cite{SeizingerKley2013} further
investigated the bouncing condition and proposed that the more realistic
condition is $\phi > 0.5$.

This critical volume filling factor for bouncing is a few times as large as
that in the laboratory experiments \citep[e.g.][]{Langkowski+etal2008}. The
origin of this discrepancy between these two approaches is not clear yet. From
a qualitative point of view, however, laboratory experiments show that fluffy
silicate aggregates with $\phi < 0.1$ tend to stick to each other
\citep{BlumWurm2000,Langkowski+etal2008,Kothe+etal2013}. This qualitative trend
is consistent with the N-body simulations. In protoplanetary discs, dust
aggregates are expected to be highly porous. In the growth of such fluffy
aggregates, the bouncing barrier would not be a strong handicap.

\section{\textbf{TOWARDS A UNIFIED MODEL}}
\label{s:unified}

Three main scenarios for the formation of planetesimals have emerged in the
last years: (1) formation via coagulation-fragmentation cycles and mass
transfer from small to large aggregates
\citep{Wurm+etal2005,Windmark+etal2012a}, (2) growth of fluffy particles and
subsequent compactification by self-gravity
\citep{Wada+etal2008,Wada+etal2009}, and (3) concentration of pebbles in the
turbulent gas and gravitational fragmentation of overdense filaments
\citep{Johansen+etal2007,Johansen+etal2009b,Kato+etal2012}.

The collision speeds at which planetesimal-sized bodies form are typically 50
m/s in models (1) and (2), as small dust aggregates are carried onto the
growing planetesimal with the sub-Keplerian wind, or even higher if the
turbulent density fluctuations are strong. During the gravitational collapse of
pebble clouds, formation mechanism (3) above, the collision speeds reach a
maximum of the escape speed of the forming body. In the cometesimal formation
zone any subsequent impacts by high-speed particles, brought in with the
sub-Keplerian flow, will at most add a few meters of compact debris to the
pebble pile over the life-time of the solar nebula (equation
\ref{eq:R2dotsimple} applied to $r=10$ AU).

This difference in impact velocity leads to substantial differences in the
tensile strengths of the planetesimals/cometesimals. Due to the relatively high
impact speeds in models (1) and (2), the growing bodies are compacted and
possess tensile strengths on the order of 1--10 kPa \citep{Blum+etal2006}.
\cite{SkorovBlum2012} pointed out that these tensile-strength values are too
high to explain the continuous dust emission of comets approaching the Sun and
favor model (3) above, for which the tensile strengths of loosely packed mm-cm
pebbles should be much smaller.

\cite{SkorovBlum2012} base their comet-nucleus model on a few assumptions: (a)
dust aggregates in the inner parts of the protoplanetary disc can only grow to
sizes of mm-cm; (b) turbulent diffusion can transport dust aggregates to the
outer disc; (c) in the outer disc, dust and ice aggregates become intermixed
with a dust-to-ice ratio found in comets; (d) cometesimals form via
gravitational instability of overdense regions of mm-cm-sized particles which
will form planetesimals with a wide spectrum of masses. While km-scale
planetesimals have not yet been observed to form in hydrodynamical simulations
of planetesimal formation, this may be an artifact of the limited numerical
resolution (see discussion in section \ref{s:collapse}).

If the comet nuclei as we find them today have been dormant in the outer
reaches of the Solar System since their formation and have not been subjected
to intense bombardment and aqueous or thermal alteration, then they today
represent the cometesimals of the formation era of the Solar System.
\cite{SkorovBlum2012} derive for their model tensile strengths of the ice-free
dusty surfaces of comet nuclei \begin{equation} T = T_0 \left(
\frac{r}{\textrm{1 mm}} \right)^{-2/3} \, , \end{equation} with $T_0=0.5$ Pa
and $r$ denoting the radius of the dust and ice aggregates composing the
cometary surface.

These extremely low tensile-strength values are, according to the
thermophysical model of \cite{SkorovBlum2012}, just sufficiently low to explain
a continuous outgassing and dust emission of comet nuclei inside a critical
distance to the Sun. This model provides strong indication that km-sized bodies
in the outer Solar System were formed by the gravitational contraction of
smaller sub-units (pebbles) at relatively low (order of m/s) velocities. Such
pebble-pile planetesimals are in broad agreement with the presence of large
quantities of relatively intact mm-sized chondrules and CAIs in chondrites from
the asteroid belt.

The cometesimals could thus represent the smallest bodies which formed by
gravitational instability of mm-cm-sized icy/rocky pebbles. These pebbles will
continue to collide inside the gravitationally collapsing clump.  Low-mass
clumps experience low collision speeds and contract as kinetic energy is
dissipated in inelastic collisions, forming small planetesimals which are made
primarily of pristine pebbles. Clumps of larger mass have higher collision
speeds between the constituent pebbles, and hence growth by
coagulation-fragmentation cycles inside massive clumps determines the further
growth towards one or more solid bodies which may go on to differentiate by
decay of short-lived radionuclides. This way coagulation is not only important
for forming pebbles that can participate in particle concentration, but
continues inside of the collapsing clumps to determine the birth size
distribution of planetesimals.


There is thus good evidence that pebbles are the primary building blocks of
planetesimals both inside the ice line (today's asteroids) and outside the ice
line (today's comets and Kuiper belt objects). This highlights the need to
understand pebble formation and dynamics better. The bouncing barrier is a
useful way to maintain a high number of relatively small pebbles in the
protoplanetary disc \citep{Zsom+etal2010}. As pebbles form throughout the
protoplanetary disc, radial drift starts typically when reaching mm-cm sizes.
Such small pebbles drift relatively fast in the outer disc but slow down
significantly (to speeds of order 10 cm/s) in the inner few AU. Pebbles with a
high ice fraction fall apart at the ice line around 3 AU, releasing refractory
grains, which can go on to form new (chondrule-like) pebbles inside the ice
line \citep{Sirono2011}, as well as water vapor which boosts formation of
large, icy pebbles outside the ice line \citep{RosJohansen2013}. Similar
release of refractories and rapid pebble growth will occur at ice lines of more
volatile species like CO \citep{Qi+etal2013}. In the optically thin very inner
parts of the protoplanetary disc, illumination by the central star leads to
photophoresis which causes chondrules to migrate outwards
\citep{Loesche+etal2013}. Radial drift of pebbles will consequently not lead to
a widespread depletion of planetesimal building blocks from the protoplanetary
disc. Instead pebble formation can be thought of as a continuous cycle of
formation, radial drift, destruction and reformation.

We propose therefore that particle growth to planetesimal sizes starts unaided
by self-gravity, but after reaching Stokes numbers roughly between 0.01 and 1
proceeds inside self-gravitating clumps of pebbles (or extremely fluffy ice
balls with similar aerodynamic stopping times). In this picture coagulation and
self-gravity are not mutually exclusive alternatives but rather two absolutely
necessary ingredients in the multifaceted planetesimal formation process.

{\bf Acknowledgments.} We thank the referee for useful comments that helped
improve the manuscript. AJ was supported by the Swedish Research Council (grant
2010-3710), the European Research Council under ERC Starting Grant agreement
278675-PEBBLE2PLANET and by the Knut and Alice Wallenberg Foundation. MB
acknowledges funding from the Danish National Research Foundation (grant number
DNRF97) and by the European Research Council under ERC Consolidator grant
agreement 616027-STARDUST2ASTEROIDS. CWO acknowledges support for this work by
NASA through Hubble Fellowship grant No.~HST-HF-51294.01-A awarded by the Space
Telescope Science Institute, which is operated by the Association of
Universities for Research in Astronomy, Inc., for NASA, under contract NAS
5-26555. HR was supported by the Swedish National Space Board (grant 74/10:2)
and the Polish National Science Center (grant 2011/01/B/ST9/05442). This work
was granted access to the HPC resources of JuRoPA/FZJ, Stokes/ICHEC and RZG
P6/RZG made available within the Distributed European Computing Initiative by
the PRACE-2IP, receiving funding from the European Community's Seventh
Framework Programme (FP7/2007-2013) under grant agreement no.~RI-283493.

\bibliographystyle{ppvi_lim1.bst}
\bibliography{bibliography}

\end{document}